\documentclass[final,3p,times,twocolumn]{elsarticle}
\usepackage{geometry}
\usepackage{mathrsfs}
\usepackage{amssymb}
\usepackage{epstopdf}

\usepackage{amssymb}
\usepackage{amsmath}
\usepackage{threeparttable}
\usepackage{subfigure}
\usepackage{graphicx}
\usepackage{caption2}
\usepackage{epstopdf}
\usepackage{algorithm} 
\usepackage{algorithmic} 
\usepackage{multirow} 
\usepackage{amsmath}
\usepackage{xcolor}
\usepackage{float}

\usepackage{amssymb}
\usepackage{amsthm}
\usepackage{setspace}
\usepackage{ragged2e}
\usepackage{float}

\usepackage{amssymb}
\usepackage{amsmath}
\usepackage{subfigure}
\usepackage{graphicx}
\usepackage{caption2}

\usepackage{algorithm} 
\usepackage{algorithmic} 
\usepackage{multirow} 
\usepackage{amsmath}
\usepackage{xcolor}
\usepackage{float}
\usepackage{epstopdf}
\usepackage{threeparttable}
\usepackage{booktabs}
\usepackage{float}
\usepackage{amssymb}

\usepackage{amssymb}
\usepackage{amsthm}
\usepackage{makecell}

\theoremstyle{definition}

\theoremstyle{remark}

\theoremstyle{problem}

\theoremstyle{property}

\usepackage{dutchcal}
\usepackage{natbib}
\usepackage{algorithm} 
\usepackage{algorithmic} 
\usepackage{multirow} 

\journal{Journal of \LaTeX\ Templates}

\begin{document}

\begin{frontmatter}

\title{Generative Graph Neural Networks for Link Prediction}

\author[tw]{Xingping~Xian}
\ead{xxp0213@gmail.com}

\author[tw]{Tao Wu\corref{mycorrespondingauthor}}
\address[tw]{School of Cybersecurity and Information Law, Chongqing University of Posts and Telecommunications, Chongqing, China.}
\ead{wutaoadeny@gmail.com}

\author[xm]{Xiaoke~Ma}
\address[xm]{School of Computer Science and Technology, XiDian University, XiAn, China.}

\author[as]{Shaojie~Qiao}
\address[as]{School of Software Engineering, Chengdu University of Information Technology, Chengdu, China.}

\author[ys]{Yabin~Shao}
\address[ys]{School of Science, Chongqing University of Posts and Telecommunications, Chongqing, China.}

\author[yw]{Chao~Wang}
\address[yw]{School of Computer and Information Science, Chongqing Normal University, Chongqing, China.}

\author[tw]{Lin~Yuan}

\author[tw]{Yu~Wu}


\cortext[mycorrespondingauthor]{Corresponding author.}

\begin{abstract}
Inferring missing links or detecting spurious ones based on observed graphs, known as link prediction, is a long-standing challenge in graph data analysis. With the recent advances in deep learning, graph neural networks have been used for link prediction and have achieved state-of-the-art performance. Nevertheless, existing methods developed for this purpose are typically discriminative, computing features of local subgraphs around two neighboring nodes and predicting potential links between them from the perspective of subgraph classification. In this formalism, the selection of enclosing subgraphs and heuristic structural features for subgraph classification significantly affects the performance of the methods. To overcome this limitation, this paper proposes a novel and radically different link prediction algorithm based on the network reconstruction theory, called GraphLP. Instead of sampling positive and negative links and heuristically computing the features of their enclosing subgraphs, GraphLP utilizes the feature learning ability of deep-learning models to automatically extract the structural patterns of graphs for link prediction under the assumption that real-world graphs are not locally isolated. Moreover, GraphLP explores high-order connectivity patterns to utilize the hierarchical organizational structures of graphs for link prediction. Our experimental results on all common benchmark datasets from different applications demonstrate that the proposed method consistently outperforms other state-of-the-art methods. Unlike the discriminative neural network models used for link prediction, GraphLP is generative, which provides a new paradigm for neural-network-based link prediction. The code is available at https://github.com/star4455/GraphLP.
\end{abstract}

\begin{keyword}
Graph Machine Learning \sep Graph Neural Networks \sep Link Prediction \sep Structural Patterns \sep Network Reconstruction.
\end{keyword}

\end{frontmatter}


\section{Introduction}

Graphs provide an elegant representation for characterizing entities and their interrelations in complex systems. Given that real-world graphs can usually only be partially observed and are often noisy, link prediction aimed at inferring missing and spurious links based on observed graphs is a paradigmatic and fundamental problem across many scientific domains, including knowledge graph completion \citep{wu2022heterogeneous}, experimental design in biological networks \citep{barzel2013network}, fake account detection in online social networks \citep{jiang2014detecting}, and product recommendation on e-commerce websites \citep{lu2012recommender}.

To address the link prediction problem, numerous heuristic methods have been proposed, including local indices such as Common Neighbors (CN) \citep{adamic2003friends}, and Resource Allocation (RA) \citep{zhou2009predicting}, global indices such as Katz \citep{katz1953new}, and SimRank \citep{jeh2002simrank}, and quasi-local indices such as the Local Path Index (LP) \citep{zhou2009predicting}. However, heuristic methods have a strong assumption on when two nodes are likely to be linked in real-world graphs and lack universal applicability to diverse areas \citep{cai2021line}. Subsequently, statistical learning-based algorithms have been proposed to obtain ground-breaking results, such as maximum likelihood-based hierarchical structure model \citep{clauset2008hierarchical}, stochastic block model \citep{guimera2009missing}, matrix factorization-based link prediction method \citep{pech2017link}, Linear Optimization (LO) link prediction method \citep{pech2019link}, and Low Frobenius norm-based Link Prediction (LFLP) method \citep{xian2020netsre}. With the proposal of network representation learning, various network embedding algorithms have been put forth so that the likelihood of a non-observed links can be estimated based on the proximity of nodes in low-dimensional vector space, including LINE \citep{tang2015line}, Node2Vec \citep{grover2016node2vec}, and DNGR \citep{cao2016deep}.

\begin{figure*}[tb]\centering
  \centering
  { \includegraphics[width = 6.86in]{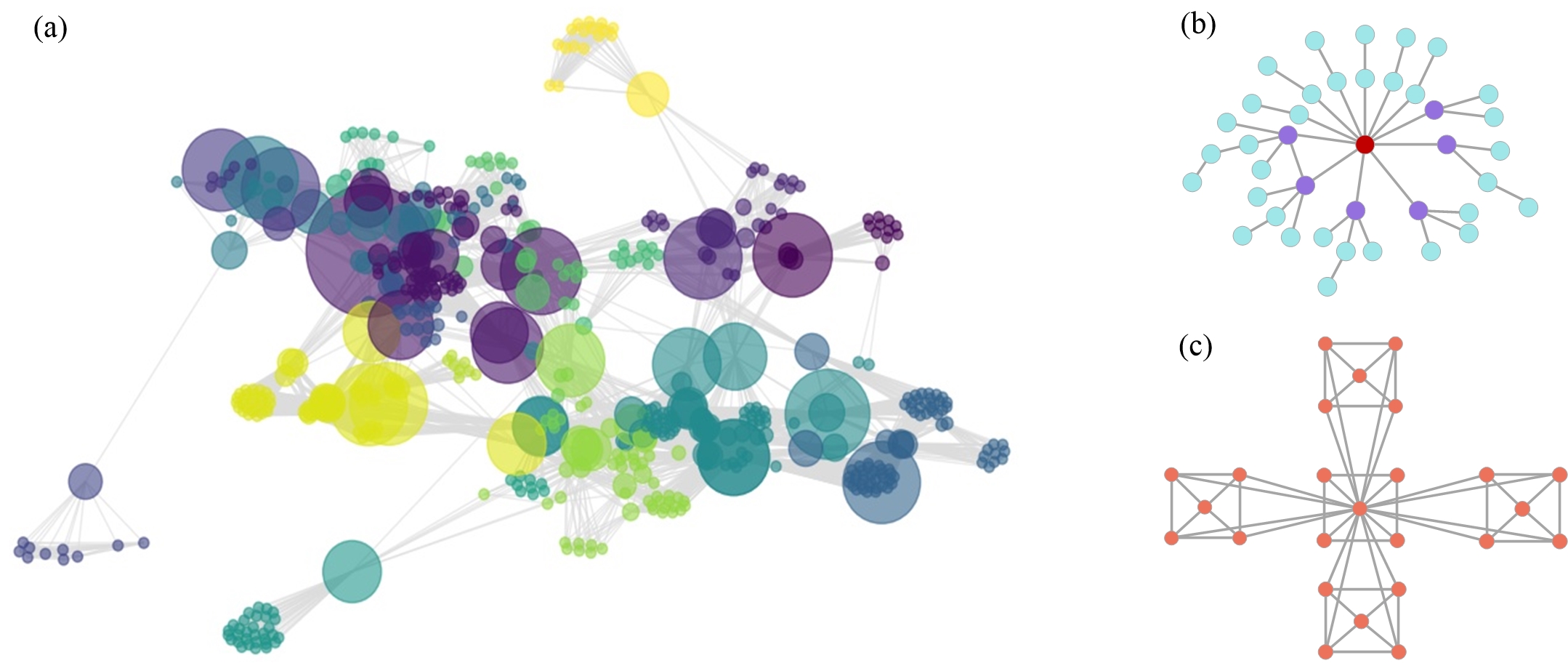}}
  \caption{An illustrative example depicting the global and high-order organizations of real-world graphs. (a) Gene network for C. elegans \citep{cho2014wormnet}. (b) Representative hierarchical star-like structure \citep{trusina2004hierarchy}. (c) Representative hierarchical modular organization \citep{ravasz2003hierarchical}. (b) and (c) depict the representative structural patterns of real-world graphs such as (a).}
  \label{fig_sim:malicuous-edge}
\end{figure*}

Recently, driven by the dramatic advances in deep learning techniques, neural networks have gradually been used to solve the link prediction problem. \cite{zhang2017weisfeiler} trained a fully-connected neural network on the enclosing subgraphs of target links for link prediction, wherein a Weisfeiler-Lehman (WL) algorithm-based graph labeling mechanism was proposed to encode subgraphs. Based on the enclosing subgraphs extracted around links, \cite{zhang2018link} trained a Graph Neural Network (GNN) for link prediction to achieve a performance comparable to that of heuristic methods. Along this line of research, \cite{pan2021neural} encoded subgraphs into random-walk transition probabilities and then computed features using these probabilities to classify positive and negative links. Although these subgraph classification-based methods have achieved state-of-the-art link prediction performance, the prediction results are found to be considerably affected by the extraction process of the $k$-hop enclosing subgraphs and the graph structure features for them. For example, in representation learning on graphs \citep{xu2018representation}, the range of enclosing subgraphs strongly depends on the graph structure, and the effective range should be different for subgraphs with varying properties.

Typically, from the perspective of subgraph classification, link prediction methods  treat subgraphs in real-world graphs independently and equivalently. That is, the global structural information of real-world graphs is totally neglected during this process. However, extensive empirical analyses indicate that real-world graphs are not locally isolated but globally relevant \citep{palla2005uncovering, wu2016integrated}; here, nodes and edges naturally portray different structural roles and contribute differently to the global organization of real-world graphs \citep{xian2020netsre, xian2021deepec}. Moreover, subgraph classification-based link prediction methods assume that real-world graphs exhibit low-order connectivity patterns and can be captured at the level of individual nodes and edges. However, empirical studies have discovered that real-world graphs exhibit high-order organizations at the level of small subgraphs, which are recursively grouped into a hierarchical structure \citep{benson2016higher, ahn2010link}. An illustrative example of the global and high-order organization in real-world graphs is depicted in Figure 1. Hence, two challenges need to be addressed for link prediction: (i) how to learn good representation preserving both local and global graph structural features? and (ii) how to characterize and utilize hierarchical structure patterns? 

To address these challenges, instead of predicting potential links through subgraph classification, this study designs a novel generative and multi-order GNN for link prediction, called GraphLP. Evidently, real-world graphs share some global properties, such as low-rank and sparsity, that can be used to provide guidance for graph learning. Hence, motivated by the network reconstruction theory \citep{guimera2009missing}, GraphLP defines a self-representation model-based collaborative inference operation to refine the observed graphs globally, which assumes that the original graph can be reconstructed utilizing the correlation between subgraph patterns. Assuming that the paths between a pair of nodes provide evidence for the existence of potential links, GraphLP extracts the local structural information via a high-order connectivity operation on the observed graphs. Thus, every neural network layer obtains the connectivity of node pairs within two-hop neighborhood, and a neural network with multiple connectivity layers captures the degree of connectivity between node pairs with various path lengths. Meanwhile, the weighted adjacency matrices generated by the connectivity operation in every neural network layer reflect the multi-order connectivity pattern in the graphs. Further, the hierarchical organizational structure of real-world graphs is explored by applying a collaborative inference operation. The contributions of this study can be summarized as follows:
\begin{itemize}
  \item \textbf{Generative framework.} Rather than subgraph classification based discriminative schemes, a novel network reconstruction-based generative GNN is proposed for link prediction, which provides a new paradigm for the application of neural networks in link prediction problem.

  \item \textbf{End-to-end learning.} Instead of designing heuristic graph structural features for subgraph representation, local and global structural patterns are extracted and fused in an end-to-end fashion for link prediction.

  \item \textbf{Algorithm.} A novel collaborative inference operation and high-order connectivity computation mechanism are developed to characterize the structural patterns in real-world graphs at different scales.


  \item \textbf{Experiment.} Extensive experiments on real-world datasets from different areas reveal that the proposed method, GraphLP, achieves promising performance and consistently outperforms other state-of-the-art methods.

\end{itemize}

\textbf{Paper Organization.} The rest of this work is organized as follows. Section 2 discusses related studies. Section 3 presents the problem definitions and describes the preliminaries. Section 4 describes the proposed method. Section 5 presents the experimental results, and finally, Section 6 presents the conclusion and discussion.

\section{Related Work}
GNNs and link prediction task have been extensively investigated in recent years. A brief review of related studies is provided in this section.



\subsection{\label{sec:level1} Graph Neural Networks }
Owing to their potential in modeling the complex structures of non-Euclidean graphs, GNNs have achieved state-of-the-art performance on almost all graph-based tasks, such as node classification, graph classification, link prediction. Based on different theories and perspectives, a plethora of different GNNs have been proposed over the years. Generally, GNNs can be divided into two categories: spectral-based and spatial-based methods. Of these, spectral-based GNNs are types of GNNs that design graph convolution operators in the spectral domain using Fourier transform. The involved convolution operation is defined as follows:
\begin{equation}
{f_1} * {f_2} = {\bf{U}}[({{\bf{U}}^{\rm{T}}}{f_1}) \odot ({{\bf{U}}^{\rm{T}}}{f_2})],
\end{equation}
where $\odot$ denotes an element-wise product. The spectral filter is defined as ${\bf{g}} = {{\bf{U}}^{\rm{T}}}{f_1}$, and the node signal ${\bf{X}}$ can be processed as follows:
\begin{equation}
{\bf{Z}} = {\bf{U}}[{\bf{g}}(\Lambda ) \odot ({{\bf{U}}^{\rm{T}}}{\bf{X}})] = {\bf{Ug}}(\Lambda ){{\bf{U}}^{\rm{T}}}{\bf{X}}.
\end{equation}
where $\bf{U}$ denotes a matrix of eigenvectors of the normalized Laplacian graph ${\bf{L}} = {\bf{I}} - {{\bf{D}}^{ - \frac{1}{2}}}{\bf{A}}{{\bf{D}}^{ - \frac{1}{2}}} = {\bf{U}}\Lambda {{\bf{U}}^{\rm{T}}}$ \citep{chen2020bridging}. Assuming that feature representation of node should be affected only by its $k$-hop neighborhood, \cite{defferrard2016convolutional} proposed a Chebyshev polynomial based $k$-localized convolution and developed a convolutional neural network, ChebNet, which eliminated the need to compute the eigenvectors of the Laplacian. Subsequently, \cite{welling2016semi} simplified the Chebshev polynomial filter using its first-order approximation and proposed the popular spectral-based method called Graph Convolutional Networks (GCNs). Notably, spatial-based GNNs define graph convolution operator based on graph topology wherein the feature vectors of node's neighbors are aggregated via a permutation-invariant function. Specifically, \cite{hamilton2017inductive} proposed a GraphSAGE approach that sampled fixed size neighborhood nodes and used max pooling, mean pooling, and LSTM pooling scheme to aggregate neighbor information. Considering the different weights of node's neighbors, \cite{velivckovic2017graph} proposed a Graph Attention Network (GAT) algorithm to calculate attention coefficient and then aggregated the neighborhood information. Other related models include PATCHY-SAN \citep{niepert2016learning}, DCNN \citep{atwood2016diffusion}, and further details on  GNNs can be found in the review \citep{zhou2020graph}.


\subsection{\label{sec:level1} Neural Networks based Link Prediction }
Following heuristic methods, matrix completion-based methods and network embedding-based methods, neural networks have been gradually applied to link prediction problem and have achieved state-of-the-art results. Specifically, \cite{zhang2017weisfeiler} proposed a link prediction method called Weisfeiler-Lehman Neural Machine (WLNM), which labeled nodes using the Weisfeiler-Lehman algorithm and encoded subgraphs to construct a feedforward neural network-based classification model. Next, from the perspective of subgraph classification, \cite{zhang2018link} proposed a novel GNN-based link prediction framework, SEAL, to learn subgraph structures and node features from local enclosing subgraphs. Along this line, to directly leverage the topology features of local subgraphs, \cite{pan2021neural} proposed a new random-walk-based pooling scheme, WalkPool, and built features for subgraph classification. Moreover, \cite{fang2021elementary} proposed a neural network-based link prediction method with only one-hop neighborhood information, which demonstrated almost equivalent performance to the WLNM and SEAL. Instead of subgraph classification, \cite{cai2021line} converted the original graph into a corresponding line graph and solved the node classification problem for link prediction. To perform link prediction for general directed or undirected complex networks, \cite{WANG2020101626} represented the adjacency matrices of networks as binary images and developed a generative adversarial networks (GANs)-based method.
In addition, because existing GNN-based methods do not scale appropriately to large graphs, \cite{louis2022sampling} extracted sparse enclosing subgraphs based on multiple random walks and presented a scalable link prediction solution, called ScaLed. To reduce the time required to determine the distances between two nodes, \cite{li2021distance} defined an anchor-based distance and proposed a new distance-enhanced GNN method for link prediction.

Among all existing methods for link prediction, the work closest to the one condidered in this study is the GANs-based method \citep{WANG2020101626}. However, this method predicts potential links via image processing within the GANs framework, whereas the proposed method conducts link prediction via GNNs-based network reconstruction. 

\subsection{\label{sec:level1} Network Structure Analysis }
Real-world graphs, also known as complex networks, are abstract representation of complex systems and have been extensively studied in the field of network science. Consequently, numerous studies have revealed that complex networks exhibit rich and diverse connectivity patterns. \cite{lu2015toward} augmented that the organization of real networks usually embodies both regularities and irregularities, where the former can be modeled and decides the extent to which the formation of a network can be explained. Notably, link predictability reflects the structural regularities in real-world networks and denotes the inherent difficulty of link prediction. \cite{xian2020netsre} proposed a self-representation network model-based method, called NetSRE, for measuring and regulating link predictability of networks. \cite{xian2021deepec} proposed a deep linear coding-based link prediction adversarial attack method by disturbing the underlying structural pattern of networks, which proved that links play global structural roles in network organization. Moreover, \cite{benson2016higher} suggested that high-order connectivity patterns are essential for understanding the fundamental structures of networks and developed a framework that identified clusters of network motifs. \cite{sales2007extracting} claimed that hierarchical structure plays an important role in complex systems. To prove the existence of hierarchical organization, an unsupervised method for extracting the hierarchical organization of complex networks was introduced and validated.

Although real-world graphs exhibit various structural patterns, most existing neural networks-based link prediction methods simply assume that they are flattened and locally isolated, and these methods judge the existence of links explicitly based only on local enclosing subgraphs. With the exception of local structural features, this study focuses on integrating global and hierarchical structural patterns into neural networks for link prediction.


\section{Problem Definition and Preliminaries}

\subsection{Problem Definition }
\textbf{Notations}. Let $\mathbcal{G = (V,E)}$ denote an undirected and unweighted graph, where $\mathbcal{V} = \{ {\mathbcal{v}_1}, \cdots ,{\mathbcal{v}_N}\}$ denotes the set of nodes and $\mathbcal{E} = \{ {\mathbcal{e}_1}, \cdots ,{\mathbcal{e}_M}\}$ denotes the set of edges. The adjacency matrix of graph $\mathbcal{G}$ is denoted as ${\bf{A}} \in {\{ 0,1\} ^{N \times N}}$, where ${{\bf{A}}_{ij}} = 1$ if nodes $i$ and $j$ are connected and ${{\bf{A}}_{ij}} = 0$ otherwise. Each edge $\mathbcal{e}$ can be represented as a node pair $({\mathbcal{u}}, {\mathbcal{v}},)$, where $\mathbcal{u}, \mathbcal{v} \in \mathbcal{V}$. Let $\mathscr{N}(u)$ denote the neighbors of node $u$, $\mathscr{N}(u) = \mathbcal{ \{ v|(u,v) \in E\} }$.

\textbf{Link Prediction}. Given an observed graph $\mathbcal{G_o = (V,E_o)}$ that corresponds to the original graph $\mathbcal{G = (V,E)}$, link prediction aims to infer the presence or absence of an edge between a pair of target nodes based on $\mathbcal{G_o}$, thereby generating a recovered graph $\mathbcal{G^ *}$ to approximate the original graph $\mathbcal{G}$. In particular, the prediction problem involves identifying a function that generates a likelihood score for a pair of nodes $(\mathbcal{u},\mathbcal{v}) \notin \mathbcal{E}$ to infer the missing link $(\mathbcal{u},\mathbcal{v})$, or to produce a likelihood score for an existing edge $(\mathbcal{u},\mathbcal{v}) \in \mathbcal{E}$ to identify spurious links. Thus, the link prediction problem can be formulated as ${s_{\mathbcal{uv}}} = f(\mathbcal{u},\mathbcal{v},{\bf{A}}|\theta )$, where $\theta$ denotes the parameter of link prediction model. In this work, ${\mathbcal{E}_m}$ and ${\mathbcal{E}_s}$ denote the identified missing and spurious links, respectively.


Note that data augmentation is a set of techniques that increases the amount and diversity of data by creating reasonable virtual data points from existing data, such that better machine learning models can be constructed based on them. According to \citep{zhou2020data}, this study considers graph data augmentation and adopts a random mapping mechanism to produce augmented graph set $\mathbcal{D}$ based on the observed graph $\mathbcal{G_o = (V,E_o)}$. Specifically, the set of all possible edges in the graph $\mathbcal{G_o}$ is denoted as $\Omega$, the existing edge set is denoted as $\mathbcal{E}_o$, and the non-existing edge set is denoted as $\mathbcal{E}_{non} = \Omega-\mathbcal{E}_o$. Thus, the candidate sets for random mapping are defined as follows:
\begin{equation}
\mathbcal{E}_{del}^c = \mathbcal{E}_o, \quad \mathbcal{E}_{add}^c = {\mathbcal{E}_{non}}.
\end{equation}
Thereafter, samples are randomly produced from the candidate sets to obtain the edge sets $\mathbcal{E}_{del}$ and $\mathbcal{E}_{add}$. Finally, a new augmented graph is generated by modifying the graph $\mathbcal{G_o}$ based on $\mathbcal{E}_{del}$ and $\mathbcal{E}_{add}$:
\begin{equation}
\mathbcal{G}' = (\mathbcal{V},(\mathbcal{E} \cup {\mathbcal{E}_{add}})\backslash {\mathbcal{E}_{del}}).
\end{equation}

Each input graph can be viewed as an instance for link prediction, owing to the generative learning scheme of the models considered in this work. Thus, the dataset containing a series of augmented graphs can be denoted as $\mathbcal{D} = \{ {\mathbcal{G}_i}|i = 1,...,t\} $ and split to yield disjoint training and validation sets. These can be denoted as ${\mathbcal{D}_{train}}$ and ${\mathbcal{D}_{val}}$ respectively, wherein the missing and spurious links of the validation set are guaranteed not to appear in the training set. The observed graph $\mathbcal{G_o}$ used to generate the augmented graphs is defined as test set ${\mathbcal{D}_{test}}$.

\subsection{Graph Convolutional Networks}
GCNs are a class of neural networks designed to generalize traditional convolution operator for non-euclidean graph-structured data. In essence, GCNs aim to learn new feature representations of nodes in graphs by exploiting their structural information.
Let adjacency matrix ${\bf{A}} \in {\{ 0,1\} ^{N \times N}}$ denote the structural information of the graph $\mathbcal{G}$, and ${\bf{X}} \in {\mathbb{R}^{N \times F}}$ denote the feature matrix of all graph nodes. Mathematically, using the output of the $l$-th layer as the input for the next layer, each neural network layer can be formulated as a nonlinear function:
\begin{equation}
{{\bf{H}}^{(l + 1)}} = f({{\bf{H}}^{(l)}},\bf{A})
\end{equation}
where ${{\bf{H}}^{(l)}}$ corresponds to the feature matrix of the $l$-th layer, and ${{\bf{H}}^{(0)}} = \bf{X}$ is the input feature matrix of the first layer. Specific GCNs models differ only in the manner in which the nonlinear function $f( \cdot )$ is instantiated. A simple example of $f( \cdot )$ is as follows:
\begin{equation}
f({{\bf{H}}^{(l)}}, {\bf{A}}) = \sigma ({\bf{A}}{{\bf{H}}^{(l)}}{{\bf{W}}^{(l)}})
\end{equation}
where $\sigma(\cdot)$ denotes a nonlinear activation function, such as a Rectified Linear Unit (ReLU), and ${{\bf{W}}^{(l)}}$ denotes a trainable weight matrix for the $l$-th layer. With this propagation rule, the neighbour's features are aggregated to represent each node at every layer, and the features become increasingly abstract by stacking layers on top of each other. However, there exist two limitations: the propagation rule simply aggregates the features of neighboring nodes but not the node itself, and the multiplication with ${\bf{A}}$ expected to change the scale of the feature vectors. That is, the nodes with a high degree will have a larger value, and the nodes with a low degree may have smaller values. To address the problems, a new propagation function, $f( \cdot )$, is presented as follows:
\begin{equation}
f({{\bf{H}}^{(l)}},{\bf{A}}) = \sigma ({{\hat {\bf{D}}}^{ - \frac{1}{2}}}\hat {\bf{A}}{{\hat {\bf{D}}}^{ - \frac{1}{2}}}{{\bf{H}}^{(l)}}{{\bf{W}}^{(l)}})
\end{equation}
where $\hat {\bf{A}}$ is obtained by adding an identity matrix ${\bf{I}}$ to the adjacency matrix $\hat {\bf{A}} = {\bf{A}} + {\bf{I}}$, ${{\hat {\bf{D}}}}$ denotes the diagonal node degree matrix of $\hat {\bf{A}}$, and ${{\hat {\bf{D}}}^{ - \frac{1}{2}}}\hat {\bf{A}}{{\hat {\bf{D}}}^{ - \frac{1}{2}}}$ denotes symmetric normalization.

\begin{table}
\begin{center}
\caption{Notations and meanings.}
\begin{tabular}{|c|l|}
\hline
Notations & Descriptions \\
\hline
$\mathbcal{G}$ &  Original graph \\
\hline
$\mathbcal{G_o}$&  Observed graph  \\
\hline
${\bf{A}}$ &  Adjacency matrix of graph \\
\hline
${\mathbcal{E}_m}$ &  Missing links \\
\hline
${\mathbcal{E}_s}$ &  Spurious links \\
\hline
$\mathbcal{D}$ & Dataset that contains augmented graphs  \\
\hline
${{\bf{H}}^{(l)}}$ & Feature matrix of $l$-th neural network layer  \\
\hline
${{\bf{W}}^{(l)}}$ & Trainable weight matrix for the $l$-th layer \\
\hline
$||\cdot|{|_0}$ & ${\ell_0}-$norm  \\
\hline
$||\cdot|{|_{2,1}}$ & ${\ell_{2,1}}-$norm  \\
\hline

\end{tabular}
\end{center}
\end{table}


\begin{figure*}[!htb]\centering
  \centering
  { \includegraphics[width = 7.0in]{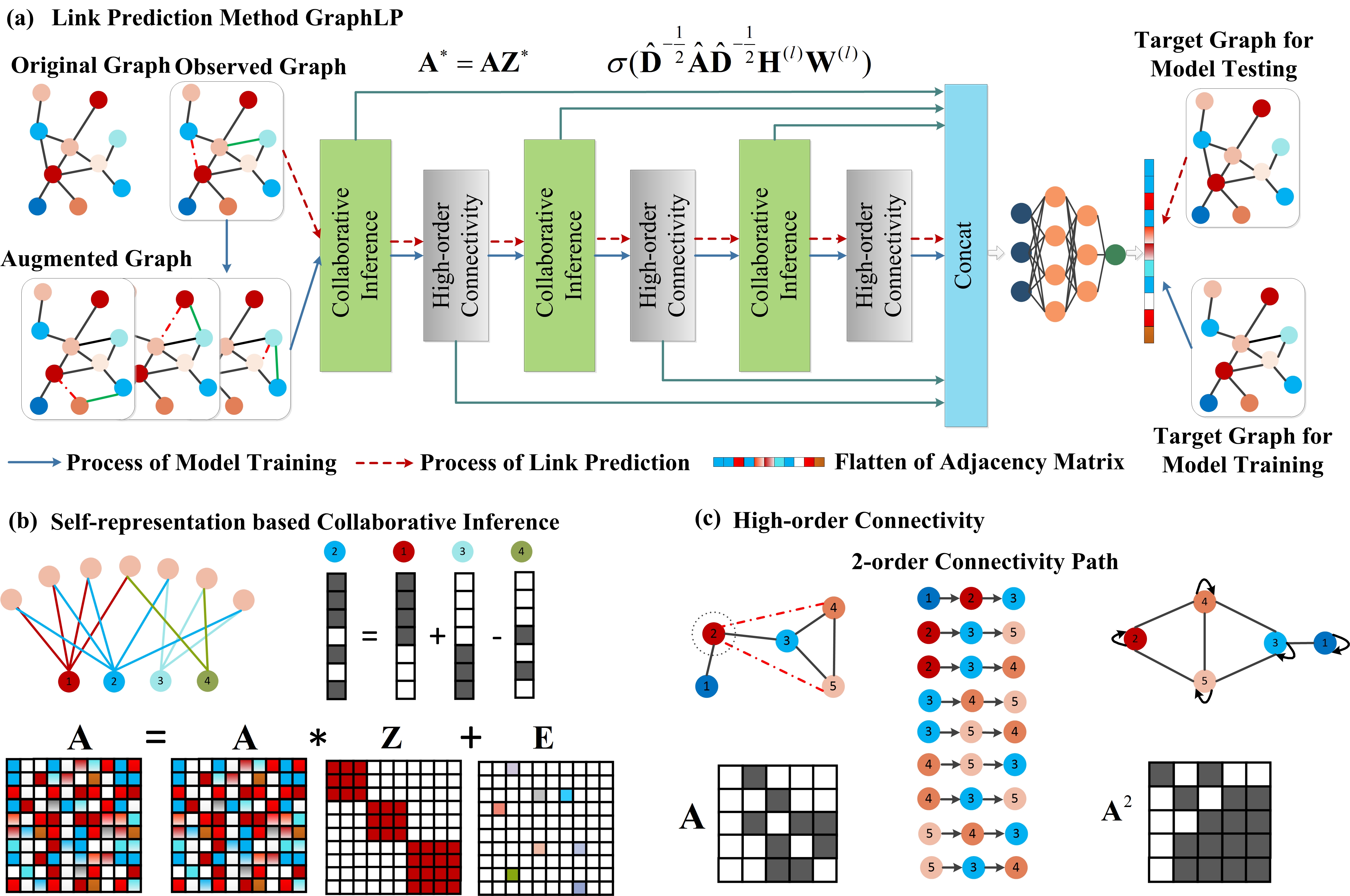}}
  \caption{Demonstration of our link prediction method, GraphLP. (a) Link prediction method, GraphLP. The original graph is perturbed using a random mapping mechanism to obtain the observed graph; after this, the observed graph is further perturbed to generate augmented graphs. These augmented graphs are fed into GraphLP to learn the model using the observed graph as the label. Subsequently, the learned model is used to infer the original graph based on the observed graph. (b) Self-representation-based collaborative inference. Based on the structural regularity of graphs, the original graph can be reconstructed by utilizing the correlation between subgraph patterns. (c) Example of high-order connectivity. In addition to the 1-hop neighborhood, multi-hop connectivity influences the existence of links. The right graph represents the two-hop connectivity of the graph on the left, and the red dotted lines in the left graph provide an example of the two-hop connectivity path of node 2. }
  \label{framework}
\end{figure*}

\subsection{Low-rank and Sparse Modeling}
Traditionally, Principal component analysis (PCA) was proposed to determine a low-dimensional representation of data while retaining as much information as possible. However, the PCA is particularly effective when dealing with Gaussian noise, which is independent and identically distributed with respect to the original data. Hence, the Robust Principal Component Analysis (RPCA) \citep{candes2011robust} has been proposed to eliminate the effect of erratic noise (outliers). PCA and RPCA methods implicitly assume that the underlying data structure is a single low-rank subspace; however, real-world data may be drawn from a union of multiple subspaces, and therefore, modeling may be inaccurate. To this end, Low-Rank Representation (LRR) \citep{liu2012robust} has been proposed.

Considering the correlation between the connectivity patterns of nodes in real-world graphs, the adjacency matrix of the graphs should be low-rank. In other words, the rows or columns of the adjacency matrix must not be linearly independent. Thus, assuming that hidden non-zero entries representing missing links can be recovered according to the adjacency matrix, \cite{pech2017link} proposed an RPCA-based link prediction method, which is formulated as the following optimization problem:
\begin{equation}
\mathop {\min }\limits_{{{\bf{X}}^ * },{\bf{E}}} {\rm{rank}}({{\bf{X}}^ * }) + {\rm{\gamma }}||{\bf{E}}|{|_0}\;s.t.,\,{\bf{A}} = {{\bf{X}}^ * } + {\bf{E}}
\end{equation}
where ${\rm{rank}}({{\bf{X}}^ * })$ denotes the rank of matrix ${{\bf{X}}^ * }$, $||\cdot|{|_0}$ is the ${\ell_0}-$norm, and $\gamma$ denots the balancing parameter. The method searches for ${{\bf{X}}^ * }$ with a low rank as low as possible and ${\bf{E}}$ as sparse as possible from $\bf{A}$. Moreover, by representing a network structure with as few representative subgraphs as possible, \cite{xian2020netsre} proposed an LRR-based link prediction method, wherein networks could be modeled via a low-rank and sparse representation, as follows:
\begin{equation}
\mathop {\min }\limits_{{\bf{Z}},{\bf{E}}} {\rm{rank}}({\bf{Z}}) + \alpha ||{\bf{Z}}|{|_0} + \beta ||{\bf{E}}|{|_{2,1}}\;s.t.,\,{\bf{A}} = {\bf{AZ}} + {\bf{E}}
\end{equation}
where ${\bf{Z}}$ denotes the representation matrix reflecting the organization principle of the network, and $||\cdot|{|_{2,1}}$ is the ${\ell_{2,1}}-$norm.

The notations used in this study are listed in Table 1.

\section{The Proposed Method}
This section presents the proposed link prediction method, GraphLP. As depicted in Figure 2, the framework of GraphLP consists of three main components:
\begin{itemize}
  \item Collaborative inference operation. There exist certain similarities between the connection patterns of individuals in a complex system such that the perturbed structure of real-world graphs can be recovered globally based on the correlation between subgraph patterns (Section 4.1).

  \item High-order connectivity computation. The existence of a link between any two target nodes is intended to be primarily determined by the connectivity degree between nodes, i.e., the number of paths and their length. Thus, the likelihood of a link can be estimated locally by computing the connectivity (Section 4.2).

  \item Pattern fusion operation. In addition to the first-order adjacency matrix, the connection patterns of nodes in the high-order adjacency matrix are also considered to be correlated, and the high-order connectivity can be reconstructed based on the collaborative inference. Thus, the graph topology can be estimated by fusing the k-order ($k \ge 1$) adjacency matrix (Section 4.3).
\end{itemize}

\subsection{Collaborative Inference Operation}
\cite{lu2015toward} suggested that link formation in real-world graphs is usually driven by both regular and irregular factors, and the former can be explained based on the mixture of multiple mechanisms, such as homophily, triadic closure, preferential attachment. Meanwhile, assuming that high-dimensional data are a mixture of simple data and are drawn from a union of multiple low-dimensional linear subspaces, the LRR has been proposed to represent the data ${\bf{A}} = [{a_1},{a_2},...,{a_N}]$ as a linear combination of the basis in a ''dictionary" ${\bf{D}} = [{d_1},{d_2},...,{d_M}]$:
\begin{equation}
\mathop {\min }\limits_{\bf{Z}} {\rm{rank}}({\bf{Z}})\;s.t.,\,{\bf{A}} = {\bf{DZ}},
\end{equation}
Thus, the optimal representation matrix ${{\bf{Z}}^ * }$ uncovers the underlying subspaces in the data. By using each subspace to model a homogeneous subset of the data, multiple subspaces in LRR can capture heterogeneous structures within the data. Therefore, considering the above ideas, the regular structure of real-world graphs can be described appropriately by the LRR model, wherein the generation mechanisms of graph organization essentially corresponds to subspaces  and the low rankness constraint captures the global correlation in graphs. Meanwhile, based on the generation mechanisms of graph organization, individual nodes may have similar connection patterns, and substructures that follow the same generation mechanism can be represented by each other, as depicted in Figure 2(b). Therefore, by using the adjacency matrix ${\bf{A}}$ as the dictionary, the real-world graph can be represented by itself, as follows:
\begin{equation}
\mathop {\min }\limits_{\bf{Z}} {\rm{rank}}({\bf{Z}})\;s.t.,\,{\bf{A}} = {\bf{AZ}}.
\end{equation}

In addition to their regular structure, real-world graphs also contain irregular components. Thus, we let matrix ${\bf{E}}$ denote such irregular connections; then, the proposed self-representation model can be modified as ${\bf{A}} = {\bf{AZ}} + {\bf{E}}$. According to the LRR, data are considered to be "sample specific", and the ${\ell_{21}}-$norm is adopted to constrain the matrix ${\bf{E}}$, i.e., $||{\bf{E}}|{|_{2,1}}$. However, although the proposed method can be used to model real-world graphs, the  low-rank model and ${\ell_{21}}-$norm constraints are usually solved using Alternating direction method (ADM), which requires a large number of iterations and has high complexity. Therefore, a reasonable strategy is to relax the constraints with Frobenius norm:
\begin{equation}
\mathop {\min }\limits_{\bf{Z}} ||{\bf{Z}}||_\mathbcal{F}^2 + \lambda ||{\bf{A}} - {\bf{AZ}}||_\mathbcal{F}^2\;s.t.\,,{\bf{A}} = {\bf{AZ}} + {\bf{E}}
\end{equation}
Let $\mathbcal{L} = ||{\bf{Z}}||_\mathbcal{F}^2 + \lambda ||{\bf{A}} - {\bf{AZ}}||_\mathbcal{F}^2$ denote the partial derivative of $\mathbcal{L}$ with respect to ${\bf{Z}}$ is ${{\partial \mathbcal{L}} \mathord{\left/ {\vphantom {{\partial \mathbcal{L}} {\partial {\bf{Z}}}}} \right.
 \kern-\nulldelimiterspace} {\partial {\bf{Z}}}} = 2{\bf{Z}} + \lambda ( - 2{{\bf{A}}^{\rm{T}}}{\bf{A}} + 2{{\bf{A}}^{\rm{T}}}{\bf{AZ}})$. By setting ${{\partial L} \mathord{\left/  {\vphantom {{\partial L} {\partial {\bf{Z}}}}} \right.  \kern-\nulldelimiterspace} {\partial {\bf{Z}}}} = 0$, the optimal representation ${{\bf{Z}}^ * }$ can be obtained as follows:
\begin{equation}
{{\bf{Z}}^ * } = \lambda {(\lambda {{\bf{A}}^{\rm{T}}}{\bf{A}}{\rm{ + }}{\bf{I}})^{ - 1}}{{\bf{A}}^{\rm{T}}}{\bf{A}}.
\end{equation}
where ${\bf{I}}$ denotes the identity matrix. Thus, in the case that the clean data is sufficient enough to represent the graph's structural patterns and the irregular connections are properly characterized, the structure perturbations can be inferred using ${\bf{A}}{{\bf{Z}}^ * }$. Hence, the collaborative inference operation is defined as follows:
\begin{equation}
\mathbcal{CI}({\bf{A}}) = \lambda {\bf{A}}{(\lambda {{\bf{A}}^{\rm{T}}}{\bf{A}}{\rm{ + }}{\bf{I}})^{ - 1}}{{\bf{A}}^{\rm{T}}}{\bf{A}}
\end{equation}

\subsection{High-order Connectivity Computation}

According to local similarity indices for link prediction, the more the number of paths two nodes possess, the greater the similarity between them. Specifically, two nodes with a high mutual connectivity are more likely to generate a link between them. Thus, n-hop-based $(n \ge 2)$ paths must be explored to characterize the local structural features for link prediction. Using a deep learning framework, the $n$-hop computation can be decomposed into two-hop operations on each neural layer. Hence, a high-order connectivity computation calculates the two hop connectivity of graph nodes in each layer, and the mutual connectivity of two nodes can be estimated by stacking the high-order connectivity computation mechanism. Assuming that the integer powers of the adjacency matrix characterizes the mutual connectivity of graph nodes, that is, ${[{{\bf{A}}^{\rm{n}}}]_{ij}}$ denotes the number of paths with length $n$ connecting nodes $i$ and $j$, the high-order connectivity computation in each neural layer can be defined based on the idea of the second power of adjacency matrix ${\bf{A}}$. From the perspective of graph convolution networks, high-order connectivity computation can be defined as
\begin{equation}
{\rm{\mathbcal{HCCA}}}({\bf{A}}) = {{{\bf{\hat D}}}^{{\rm{ - }}\frac{1}{2}}}{\bf{\hat A}}{{{\bf{\hat D}}}^{{\rm{ - }}\frac{1}{2}}}\mathbcal{CI}({\bf{A}}),
\end{equation}
where the weighted adjacency matrix generated by the proposed collaborative inference operation is viewed as the features of graph nodes. Figure 3 illustrates a high-order connectivity computation. As presented in Equation (15), the global and local structural features can be captured for link prediction at the level of individual nodes and edges. Thus, the nonlinear propagation function can be defined as follows:
\begin{equation}
{{\bf{H}}^{(l + 1)}} =  {{{\bf{\hat D}}}^{{\rm{ - }}\frac{1}{2}}}{\bf{\hat A}}{{{\bf{\hat D}}}^{{\rm{ - }}\frac{1}{2}}}\mathbcal{CI}({{\bf{H}}^{(l)}}){{\bf{W}}^{(l)}}.
\end{equation}
Thus, the hierarchical structure of real-world graphs can be characterized by executing the nonlinear propagation function iteratively, in which ${\rm{\mathbcal{HCCA}}}({{\bf{H}}^{(l)}}) = {{{\bf{\hat D}}}^{{\rm{ - }}\frac{1}{2}}}{\bf{\hat A}}{{{\bf{\hat D}}}^{{\rm{ - }}\frac{1}{2}}}\mathbcal{CI}({{\bf{H}}^{(l)}})$ represents the high-order connectivity of graph nodes, as depicted in Figure 2(c), and $\mathbcal{CI}({{\bf{H}}^{(l + 1)}})$ denotes the collaborative inference.

\begin{figure}[!tb]\centering
  \centering
  { \includegraphics[width = 3.5in]{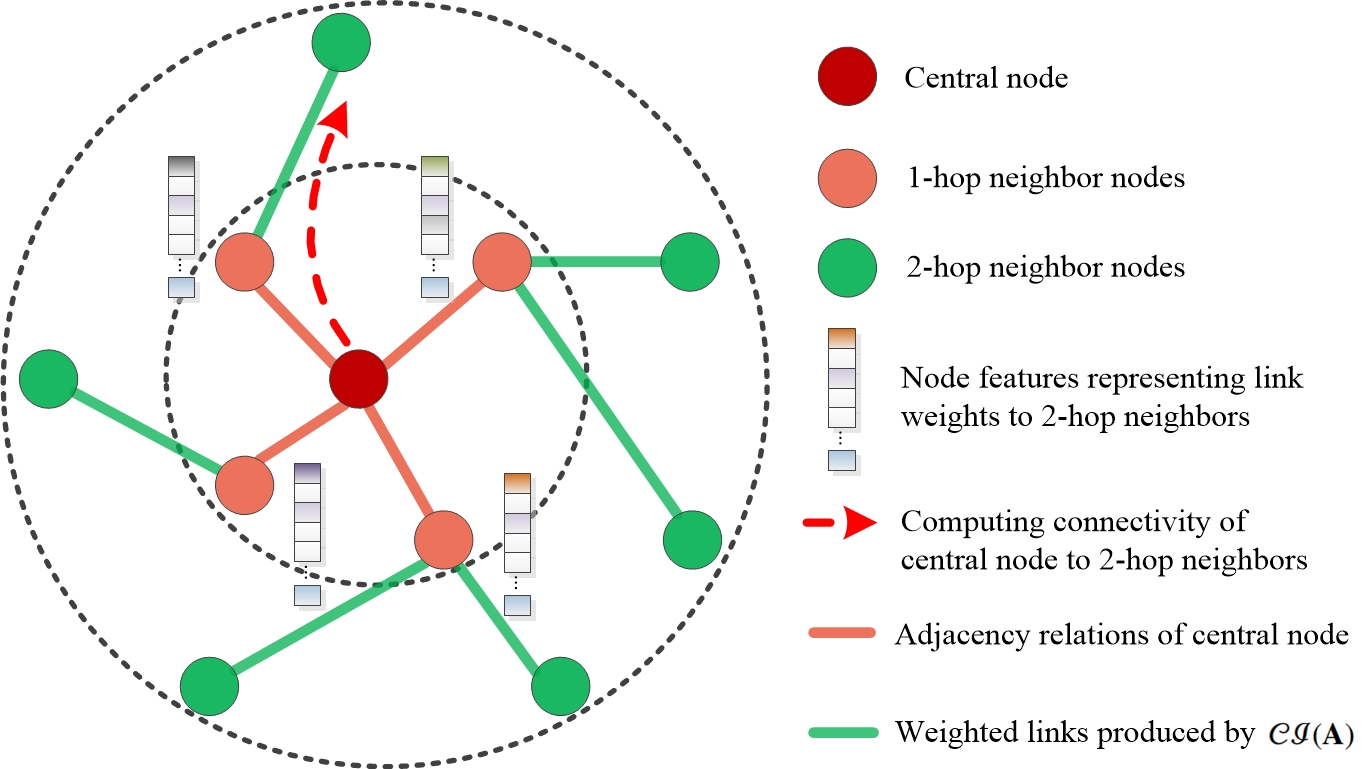}}
  \caption{Illustration of high-order connectivity computation.}
  \label{framework}
\end{figure}

%

\subsection{Pattern Fusion Operation}

To estimate the likelihood of potential links, the output of the $(l-1)$-th layer, i.e., ${{\bf{H}}^{(l)}}$, is fed as the input of the $l$-th layer.
Based on $\mathbcal{CI}({{\bf{H}}^{(l)}})$ and ${\rm{\mathbcal{HCCA}}}({{\bf{H}}^{(l)}})$, the shallow layers extract the low-order global and local structure features, while the deep layers extract the high-order global and local structure features. Meanwhile, the effective range that the local structure features drawn from increases as the model depth increases. Therefore, the structure features in different range at various order, i.e., ${\rm{\mathbcal{HCCA}}}({{\bf{H}}^{(l)}})$ and $\mathbcal{CI}({{\bf{H}}^{(l)}})$, $0 \le l \le L$, all contribute to the inference of potential links, although the exact extent of their contribution depends on the graph data.

To overcome the issues mentioned above, in addition to being used as the inputs of the next layer, the outputs of neural network layers are mapped to skip a block of several layers based on residual connections, as illustrated in Figure 2(a). Next, all outputs are concatenated and used as the input of a two-layer Multi-layer Perceptron (MLP), which is defined as:
\begin{equation}
{\bf{O}}{\rm{ = MLP(concat(}}\mathbcal{CI}({{\bf{H}}^{(l)}}),{\rm{\mathbcal{HCCA}}}({{\bf{H}}^{(l)}}){\rm{)),}}0 \le l \le L.
\end{equation}
where ${\bf{O}}$ is a vector containing the probabilities of links between all possible node pairs, and missing and spurious links can be inferred based on it.


\subsection{Model Training}
To train the proposed model, augmented graphs generated based on the observed graph are used as training data, and the adjacency matrix of the observed graph is flatten as its labels ${\bf{Y}}$, where ${{{\bf{Y}}_{i * N + j}}}$ denotes the existence of the link between nodes $i$ and $j$. Correspondingly, ${\bf{O}}$ represents the prediction results  obtained by the proposed model for all possible links. Here, the Binary Cross-Entropy (BCE) is used as the loss function:
\begin{equation}
\mathbcal{L} =  - \frac{1}{{{N^2}}}\sum\limits_{i = 1}^{{N^2}} {{{\bf{Y}}_i}\log ({{\bf{O}}_i}) + (1 - {{\bf{Y}}_i})\log (1 - {{\bf{O}}_i})}.
\end{equation}
The learned model is then deployed on the observed graph to reconstruct the original graph. The training process of GraphLP is outlined in Algorithm 1.

\renewcommand{\algorithmicrequire}{\textbf{Input:}}
\renewcommand{\algorithmicensure}{\textbf{Output:}}
\begin{algorithm}[!htb]
\caption{Training Process of GraphLP}
\label{alg::conjugateGradient}
\begin{algorithmic}[1]
\REQUIRE Training set ${\mathbcal{D}_{train}}$, validation set ${\mathbcal{D}_{val}}$, and test set ${\mathbcal{D}_{test}}$, number of neural network layers $L$.
\ENSURE The well-trained model GraphLP.

\WHILE {not convergence}
    \FOR{$0 \le l \le L$}
        \STATE Conduct collaborative inference operation using (14);
        \STATE Compute high-order connectivity using (16);
    \ENDFOR

\STATE Fuse the outputs based on MLP using (17);

\STATE Update the model by minimizing the loss function (18);

\ENDWHILE

\end{algorithmic}
\end{algorithm}

\subsection{Model Analysis}
(1) Generalized local similarity indices. The high-order connectivity computation ${\rm{\mathbcal{HCCA}}}({{\bf{H}}^{(l)}})$ in every neural network layer is essentially the second power of the adjacency matrix, and it obtains the connectivity of node pairs within two-hop neighborhood. As the model depth increases, the connectivity of node pairs in a wider range is considered. Thus, GraphLP can degenerate to ${S_{ij}} = {{\bf{A}}^2} + \alpha {{\bf{A}}^3} + \beta {{\bf{A}}^4} +  \cdot  \cdot  \cdot $ when collaborative inference and deep learning mechanism are abolished.

(2) Connection to WalkPool. WalkPool \citep{pan2021neural} first generates node representations based on GNN and encodes them into edge weights of the extracted enclosing subgraphs; following this, it uses the edge weights to compute the transition probabilities of random walk. Next, the method calculates a list of features based on the transition probabilities to classify the subgraphs. However, for an enclosing subgraph $\mathbcal{G} = (\mathbcal{V},\mathbcal{E})$, its variants ${\mathbcal{G}^ + } = (\mathbcal{V},\mathbcal{E} \cup \{ i,j\} )$ and ${\mathbcal{G}^ - } = (\mathbcal{V},\mathbcal{E} \backslash \{ i,j\} )$ are used as positive and negative samples, respectively. In essence, this method discriminates only those subgraphs that differ by a single edge and is not suitable for practical link prediction scenarios. In contrast, GraphLP can predict any potential links based on graph structure features.

(3) Connection to LFLP. The LFLP \citep{xian2020netsre} constructs an adjacency matrix based on a self-representation model and then combines it with the observed network to identify missing and spurious links. The collaborative inference operation $\mathbcal{CI}({{\bf{H}}^{(l)}})$ of our work is similar to that in the LFLP with respect to modeling the global structure of graphs; however, the difference is that only low-order global structural features are considered in LFLP, whereas multi-order global and local structural features are characterized based on the deep-learning framework in GraphLP.

\section{Experiments}
Further, extensive experiments are conducted on real-world graphs to evaluate the performance of the proposed method GraphLP: (1) Compare GraphLP with state-of-the-art methods; (2) Compare GraphLP with traditional baseline methods; (3) Model architecture analysis; (4) Model sensitivity analysis. Here, Area Under Curve (AUC) and Average Precision (AP) are adopted to evaluate the performance of the methods. Furthermore, Precision is used to verify the superiority of GraphLP over traditional link prediction methods. Based on the link prediction results ${\bf{O}}$, the scores are sorted in descending and ascending orders, and following this, their top-L links are taken as the predicted missing and spurious links. Note that Precision is defined by calculating the ratio of accurately discovered links to the total number of links in the probe set:

\begin{equation}
{{{\rm{Precision = \mathbcal{T}}}} \mathord{\left/  {\vphantom {{{\rm{Precision = \mathbcal{T} }}} {\mathbcal{R}}}} \right. \kern-\nulldelimiterspace} {\mathbcal{R}}}
\end{equation}
where $\mathbcal{T}$ is the number of accurately identified links, and $\mathbcal{R}$ is the total number of links in the probe set.

\subsection{\label{sec:level1}Experimental Settings}

\subsubsection{\label{sec:level1} Experimental Datasets}
Herein, seven widely used graph datasets are used for link prediction. (1) USAir ~\citep{rossi2015network}. This is the transportation network of the United States, including 332 airports as nodes and 2,126 airlines as edges, connecting the United States worldwide. The average node degree is 12.81. (2) C.ele ~\citep{ watts1998collective}. This is a neural network of C. elegans, with 297 neurons representing nodes and 2,148 synaptic connections representing edges. The average node degree is 14.46. (3) PB ~\citep{ackland2005mapping}. This dataset is a network of hyperlinks between weblogs on US political blogs, with 1,222 blogs on US politics as nodes and 16,714 hyperlinks between blogs as edges. The average node degree is 27.36. (4) NS ~\citep{newman2006finding}. This is an undirected co-authorship network with 1,589 nodes and 2,742 edges, where the nodes denote the scientists engaged in network science research, and the edges denote two scientists have co-authored a publication. The average node degree is 3.45. (5) Yeast ~\citep{von2002comparative}. This represents a protein-protein interaction network formed in yeast with 2,375 proteins as nodes and 11,693 protein-protein interactions as edges. The average node degree is 9.85. (6) E.coli ~\citep{zhang2018beyond}. This is a pairwise reaction network of metabolites with 1,805 nodes and 14,660 edges. The average node degree is 12.55. (7) Router \citep{spring2002measuring}. It is a snapshot of the Internet structure at the level of autonomous systems, with 5,022 nodes and 6,258 edges, in which the nodes represent routers and the edges represent the data transmission between routers. The average node degree is 2.49. The properties of the datasets are listed in Table 2.

To extensively validate the performance of the proposed method, $90\%$ and $50\%$ of the links of the original graph are selected randomly to first construct the observed graphs. Thereafter, based on the observed graph $\mathbcal{G_o}$, $10\%$ nonexisting links are add randomly as spurious links, and $10\%$ existing links are removed randomly as missing links, denoted as $\mathbcal{E}_{del}$ and $\mathbcal{E}_{add}$ respectively, to generate the augmented graph set $\mathbcal{D} = \{ {\mathbcal{G}_i}|i = 1,...,t\} $. Following this, $90\%$ and $10\%$ graphs are randomly select from $\mathbcal{D}$ as the training and validation set, respectively, and the observed graph $\mathbcal{G_o}$ is used as the test set.

\begin{table}[h]
	\setlength{\tabcolsep}{1.3mm}
	\centering
	\caption{Summary of the datasets. ACC is the average clustering coefficient, and AD is the average node degree.}
    \begin{tabular}{c|cccccccc}
      \hline
      \multirow{1}{*}{Dataset} & \multirow{1}{*}{USAir} & \multirow{1}{*}{NS}  & \multirow{1}{*}{PB} & \multirow{1}{*}{Yeast} & \multirow{1}{*}{C.ele} & \multirow{1}{*}{Router}  & \multirow{1}{*}{E.coli}\\ \hline
      \multirow{1}{*}{Node}   & 332  & 1589 & 1222 & 2375 & 297 & 5022 & 1805\\  \hline

      \multirow{1}{*}{Edges} & 2126  & 2742  & 16714 & 11693 & 2148 & 6258 & 14660 \\  \hline

      \multirow{1}{*}{ACC} & 0.625  & 0.638  & 0.320 & 0.306 & 0.292 & 0.012 & 0.516 \\     \hline
      \multirow{1}{*}{AD} & 12.81  & 3.45  & 27.36  & 9.85 & 14.46 & 2.49 & 12.55 \\    \hline
	\end{tabular}%
	\label{tab:time-result}%
\end{table}%

\subsubsection{\label{sec:level1} Comparison Methods}

 The proposed method was compared with six state-of-the-art deep learning-based link prediction methods, including:

(1) Weisfeiler-Lehman graph kernel (WLK) ~\citep{shervashidze2011weisfeiler} is a fast feature extraction scheme based on the WL test for graph isomorphism, which maps the original graph to a graph sequence and adds the pair-wise similarities between the graphs.

(2) Weifeiler-Lehmam Neural Machine (WLNM) ~\citep{zhang2017weisfeiler} is a subgraph classification-based link prediction method that leverage deep learning to automatically learn topological features from enclosing subgraphs.

(3) Node2Vec ~\citep{grover2016node2vec} is a network embedding method that encodes proximity information into low-dimensional vectors. The node features and low-dimensional vectors are then fed into the MLP for link prediction.

(4) LINE ~\citep{tang2015line} learns network embeddings that preserve the first-order and second-order proximity, and the resulting low-dimensional vectors are used for link prediction.


(5) SEAL ~\citep{zhang2018link} extracts the enclosing subgraphs of positive and negative links and marks different roles of their nodes. The method then trains a GNN based on the node information matrix to classify subgraphs for link prediction.

(6) WalkPool (WP) ~\citep{pan2021neural} is a subgraph classification-based link prediction method. It encodes node feature and graph topology into the transition probabilities of random walk, and following this, a list of features is computed to classify subgraphs.


\subsubsection{\label{sec:level1} Parameter Settings}
GraphLP is implemented on a Pytorch platform with a NVIDIA GeForce RTX GPU and optimized using Adam optimizer. All models are implemented using Python 3.6. The learning rate is set to 0.0012 for the NS dataset and 0.0005 for the other graphs. For all the datasets, the weight decay is set to 0.0. The number of epochs on the E.coli and Yeast dataset is 300, whereas it was 200 on the other datasets. Dropout is applied to the MLP, and the dropout rate is set to 0.5 on Router and 0.2 on the others. The trade-off parameter $\lambda$ is set to 0.13, and the number of neural network layers in the GraphLP is set to three. The detailed hyperparameter settings for the model are listed in Table 3.

\begin{table}[h]
	\setlength{\tabcolsep}{1.3mm}
	\centering
	\caption{Hyperparameter setting for the proposed method.}
    \begin{tabular}{c|c}
      \hline
      \multirow{1}{*}{Name} & \multirow{1}{*}{Value} \\ \hline
      \multirow{1}{*}{optimizer}   & {Adam} \\  \hline
      \multirow{1}{*}{loss function}   & {binary\_cross\_entropy} \\  \hline
      \multirow{1}{*}{learning rate} & {NS=0.0012, others=0.0005} \\  \hline
      \multirow{1}{*}{weight decay} & 0.0 \\     \hline
      \multirow{1}{*}{epochs} & Ecoli, Yeast=300, others=200 \\    \hline
      \multirow{1}{*}{dropout} & {Router=0.5, others=0.2} \\    \hline
      \multirow{1}{*}{$\lambda$} & 0.13 \\    \hline
      \multirow{1}{*}{number of network layers} & 3 \\    \hline
	\end{tabular}%
	\label{tab:time-result}%
\end{table}%


\begin{table*}[htbp]\centering
	\setlength{\tabcolsep}{1.8mm}
	\centering
	\caption{Prediction measured by AUC (90$\%$ observed links). \textbf{Bold} numbers are the best results of all methods.}
    \begin{tabular}{c|cccccccc}
      \hline
      \multirow{1}{*}{Data} & \multirow{1}{*}{USAir} & \multirow{1}{*}{NS}  & \multirow{1}{*}{PB} & \multirow{1}{*}{Yeast} & \multirow{1}{*}{C.ele} & \multirow{1}{*}{Router} & \multirow{1}{*}{E.coli} \\ \hline

      \multirow{1}{*}{WLK} & 96.63 $\pm$ 0.73 & 98.57 $\pm$ 0.51  & 93.83 $\pm$ 0.59 & 95.86 $\pm$ 0.54  & 89.72 $\pm$ 1.67 & 87.42 $\pm$ 2.08 & 96.94 $\pm$ 0.29 \\ \hline

      \multirow{1}{*}{WLNM} & 95.95 $\pm$ 1.10 & 98.61 $\pm$ 0.49  & 93.49 $\pm$ 0.47 & 95.62 $\pm$ 0.52 & 86.18 $\pm$ 1.72 & 94.41 $\pm$ 0.88 & 97.21 $\pm$ 0.27  \\ \hline

      \multirow{1}{*}{Node2Vec} & 91.44 $\pm$ 1.78 & 91.52 $\pm$ 1.28  & 85.79 $\pm$ 0.78 & 93.67 $\pm$ 0.46 & 84.11 $\pm$ 1.27 & 65.46 $\pm$ 0.86 & 90.82 $\pm$ 1.49  \\ \hline

      \multirow{1}{*}{LINE} & 81.47 $\pm$ 10.71 & 80.63 $\pm$ 1.90  & 76.95 $\pm$ 2.76 & 87.45 $\pm$ 3.33 & 69.21 $\pm$ 3.14 & 67.15 $\pm$ 2.10 & 82.38 $\pm$ 2.19  \\ \hline

      \multirow{1}{*}{SEAL} & 97.09 $\pm$ 0.7 & 98.85 $\pm$ 0.41  & 95.01 $\pm$ 0.34 & 97.91 $\pm$ 0.52 & 90.30 $\pm$ 1.35 & 96.38 $\pm$ 1.45 & 97.64 $\pm$ 0.22  \\ \hline

      \multirow{1}{*}{WP} & 98.68 $\pm$ 0.48 & 98.95 $\pm$ 0.41  & 95.60 $\pm$ 0.37 & 98.37 $\pm$ 0.25 & 95.79 $\pm$ 1.09 & 97.27 $\pm$ 0.28 & 98.58 $\pm$ 0.19  \\ \hline

      \multirow{1}{*}{GraphLP} & \textbf{99.26 $\pm$ 1.01} & \textbf{99.64 $\pm$ 0.98}  & \textbf{99.73 $\pm$ 0.25} & \textbf{99.41 $\pm$ 0.15} & \textbf{99.90 $\pm$ 0.14} & \textbf{99.02 $\pm$ 0.19} & \textbf{99.23 $\pm$ 0.23}  \\ \hline

	\end{tabular}%
	\label{tab:time-result}%
\end{table*}%

\subsection{\label{sec:level1} Experimental Result}
For 90$\%$ of the observed links, the results about the AUC and AP with standard deviations are presented in Table 4 and 5, which indicate that GraphLP significantly outperforms other state-of-the-art algorithms in terms of both AUC and AP, with exception of the NS and Router datasets. The results demonstrate that the learning of local and global graph structure entirely characterizes the underlying structural patterns; thus, the missing links and spurious links can be better identified. Table 4 indicates that GraphLP significantly improves the AUC on the PB, C.ele, and Router datasets, with approximately 4$\%$, 7$\%$, and 3$\%$ performance improvement,  respectively, compared to the WP algorithm. In addition, the proposed method still performs better than other state-of-the-art methods on the USAir, Yeast and NS datasets. Moreover, the results for the AP presented in Table 5 also indicate that GraphLP outperforms state-of-the-art methods on most of datasets, and GraphLP achieves a maximum performance enhancement of approximately 9$\%$ compared to the best performing graph neural network method WP.

For 50$\%$ of the observed links, the results also demonstrate that the proposed model achieves remarkable performance compared to the methods, as described in Table 6 and 7. The results illustrate that, as the amount of structure perturbation increases, GraphLP can still appropriately learn the real graph structure, thus recovering the original graph effectively. Therefore, the values of the AUC and AP decreased to a lower extent. Furthermore, by comparing Table 4 with Table 6 and Table 5 with Table 7, we can infer that the AUC and AP values drop faster for the other state-of-the-art methods than those for GraphLP, which demonstrates that GraphLP can better capture the underlying structural patterns to demonstrate better performance.

\begin{table*}[htbp]\centering
	\setlength{\tabcolsep}{1.8mm}
	\centering
	\caption{Prediction measured by AP (90$\%$ observed links). \textbf{Bold} numbers are the best results of all methods.}
    \begin{tabular}{c|cccccccc}
      \hline
      \multirow{1}{*}{Data} & \multirow{1}{*}{USAir} & \multirow{1}{*}{NS}  & \multirow{1}{*}{PB} & \multirow{1}{*}{Yeast} & \multirow{1}{*}{C.ele} & \multirow{1}{*}{Router} & \multirow{1}{*}{E.coli} \\ \hline

      \multirow{1}{*}{WLK} & 96.82 $\pm$ 0.84 & 98.79 $\pm$ 0.40  & 93.34 $\pm$ 0.89 & 96.82 $\pm$ 0.35  & 88.96 $\pm$ 2.06 & 86.59 $\pm$ 2.23 & 97.25 $\pm$ 0.42 \\ \hline

      \multirow{1}{*}{WLNM} & 95.95 $\pm$ 1.13 & 98.81 $\pm$ 0.49  & 92.69 $\pm$ 0.64 & 96.40 $\pm$ 0.38 & 85.08 $\pm$ 2.05 & 93.53 $\pm$ 1.09 & 97.50 $\pm$ 0.23  \\ \hline

      \multirow{1}{*}{Node2Vec} & 89.71 $\pm$ 2.97 & 94.28 $\pm$ 0.91  & 84.79 $\pm$ 1.03 & 94.90 $\pm$ 0.38 & 83.12 $\pm$ 1.90 & 68.66 $\pm$ 1.49 & 90.87 $\pm$ 1.48  \\ \hline

      \multirow{1}{*}{LINE} & 97.70 $\pm$ 11.76 & 85.17 $\pm$ 1.65  & 78.82 $\pm$ 2.71 & 90.55 $\pm$ 2.39 & 67.51 $\pm$ 2.72 & 71.92 $\pm$ 1.53 & 86.45 $\pm$ 1.82  \\ \hline

      \multirow{1}{*}{SEAL} & 97.13 $\pm$ 0.80 & 99.06 $\pm$ 0.37  & 94.55 $\pm$ 0.43 & 98.33 $\pm$ 0.37 & 89.48 $\pm$ 1.85 & 96.23 $\pm$ 1.71 & 98.03 $\pm$ 0.20  \\ \hline

      \multirow{1}{*}{WP} & 98.66 $\pm$ 0.55 & \textbf{99.09 $\pm$ 0.29}  & 95.28 $\pm$ 0.41 & 98.64 $\pm$ 0.28 & 91.53 $\pm$ 1.33 & \textbf{97.20 $\pm$ 0.38} & 98.79 $\pm$ 0.21  \\ \hline

      \multirow{1}{*}{GraphLP} & \textbf{99.91 $\pm$ 1.03} & 98.94 $\pm$ 0.96  & \textbf{98.32 $\pm$ 1.43} & \textbf{98.74 $\pm$ 0.16} & \textbf{99.41 $\pm$ 0.42} & 79.30 $\pm$ 0.19 & \textbf{98.96 $\pm$ 0.19}  \\ \hline

	\end{tabular}%
	\label{tab:time-result}%
\end{table*}%

\begin{table*}[htbp]
	\setlength{\tabcolsep}{1.8mm}
	\centering
	\caption{Prediction measured by AUC ( 50$\%$ observed links). \textbf{Bold} numbers are the best results of all methods.}
    \begin{tabular}{c|cccccccc}
      \hline
      \multirow{1}{*}{Data} & \multirow{1}{*}{USAir} & \multirow{1}{*}{NS}  & \multirow{1}{*}{PB} & \multirow{1}{*}{Yeast} & \multirow{1}{*}{C.ele} & \multirow{1}{*}{Router} & \multirow{1}{*}{E.coli} \\ \hline

      \multirow{1}{*}{WLK} & 91.93 $\pm$ 0.71 & 87.27 $\pm$ 1.71  & 92.54 $\pm$ 0.33 & 91.15 $\pm$ 0.35  & 83.29 $\pm$ 0.89 & 71.25 $\pm$ 4.37 & 92.38 $\pm$ 0.46 \\ \hline

      \multirow{1}{*}{WLNM} & 91.42 $\pm$ 0.95 & 87.61 $\pm$ 1.63  & 90.93 $\pm$ 0.23 & 92.22 $\pm$ 0.32 & 75.72 $\pm$ 1.33 & 86.10 $\pm$ 0.52 & 92.81 $\pm$ 0.30  \\ \hline

      \multirow{1}{*}{Node2Vec} & 84.63 $\pm$ 1.58 & 80.29 $\pm$ 1.20  & 79.29 $\pm$ 0.67 & 90.18 $\pm$ 0.17 & 75.53 $\pm$ 1.23 & 62.45 $\pm$ 0.81 & 84.73 $\pm$ 0.81  \\ \hline

      \multirow{1}{*}{LINE} & 72.51 $\pm$ 12.19 & 65.96 $\pm$ 1.60  & 75.53 $\pm$ 1.78 & 79.44 $\pm$ 7.90 & 59.46 $\pm$ 7.08 & 62.43 $\pm$ 3.10 & 74.50 $\pm$ 11.10  \\ \hline

      \multirow{1}{*}{SEAL} & 93.36 $\pm$ 0.67 & 90.88 $\pm$ 1.18  & 93.79 $\pm$ 0.25 & 93.90 $\pm$ 0.54 & 82.33 $\pm$ 2.31 & 86.64 $\pm$ 1.58 & 94.18 $\pm$ 0.41  \\ \hline

      \multirow{1}{*}{WP} & 95.50 $\pm$ 0.74 & 90.97 $\pm$ 0.96  & 94.57 $\pm$ 0.16 & 95.00 $\pm$ 0.21 & 87.62 $\pm$ 1.39 & 88.13 $\pm$ 0.61 & 95.33 $\pm$ 0.30  \\ \hline

      \multirow{1}{*}{GraphLP} & \textbf{98.97 $\pm$ 0.15} &  \textbf{97.08 $\pm$ 0.14}  & \textbf{98.19 $\pm$ 0.10} & \textbf{98.74 $\pm$ 0.25} & \textbf{97.96 $\pm$ 0.14} & \textbf{98.10 $\pm$ 0.15} &  \textbf{98.05 $\pm$ 0.14}  \\ \hline

	\end{tabular}%
	\label{tab:time-result}%
\end{table*}%

\begin{table*}[htbp]
	\setlength{\tabcolsep}{1.8mm}
	\centering
	\caption{Prediction measured by AP ( 50$\%$ observed links). \textbf{Bold} numbers are the best results of all methods.}
    \begin{tabular}{c|cccccccc}
      \hline
      \multirow{1}{*}{Data} & \multirow{1}{*}{USAir} & \multirow{1}{*}{NS}  & \multirow{1}{*}{PB} & \multirow{1}{*}{Yeast} & \multirow{1}{*}{C.ele} & \multirow{1}{*}{Router} & \multirow{1}{*}{E.coli} \\ \hline

      \multirow{1}{*}{WLK} & 93.34 $\pm$ 0.51 & 89.97 $\pm$ 1.02  & 92.34 $\pm$ 0.34 & 93.55 $\pm$ 0.46  & 83.20 $\pm$ 0.90 & 75.49 $\pm$ 3.43 & 94.51 $\pm$ 0.32 \\ \hline

      \multirow{1}{*}{WLNM} & 92.54 $\pm$ 0.81 & 90.10 $\pm$ 1.11  & 91.01 $\pm$ 0.20 & 93.93 $\pm$ 0.20 & 76.12 $\pm$ 1.08 & 86.12 $\pm$ 0.68 & 94.47 $\pm$ 0.21  \\ \hline

      \multirow{1}{*}{Node2Vec} & 82.51 $\pm$ 2.08 & 86.01 $\pm$ 0.87  & 77.21 $\pm$ 0.97 & 92.45 $\pm$ 0.23 & 72.91 $\pm$ 1.74 & 66.77 $\pm$ 0.57 & 85.41 $\pm$ 0.94  \\ \hline

      \multirow{1}{*}{LINE} & 71.75 $\pm$ 11.85 & 71.53 $\pm$ 0.97  & 78.72 $\pm$ 1.24 & 83.06 $\pm$ 9.70 & 60.71 $\pm$ 6.26 & 64.87 $\pm$ 6.76 & 75.98 $\pm$ 14.45  \\ \hline

      \multirow{1}{*}{SEAL} & 94.15 $\pm$ 0.54 & 92.21 $\pm$ 0.97  & 93.42 $\pm$ 0.19 & 95.32 $\pm$ 0.38 & 81.99 $\pm$ 2.18 & 87.79 $\pm$ 1.71 & 95.67 $\pm$ 0.24  \\ \hline

      \multirow{1}{*}{WP} & 95.87 $\pm$ 0.74 & 92.33 $\pm$ 0.76  & 94.22 $\pm$ 0.27 & 96.15 $\pm$ 0.13 & 86.25 $\pm$ 1.42 & \textbf{89.17 $\pm$ 0.55} &  96.36 $\pm$ 0.34  \\ \hline

      \multirow{1}{*}{GraphLP} & \textbf{97.96 $\pm$ 0.09} &  \textbf{93.08 $\pm$ 0.08}  & \textbf{96.27 $\pm$ 0.10} & \textbf{97.27 $\pm$ 0.09} & \textbf{95.89 $\pm$ 0.11} & 79.23 $\pm$ 0.14 & \textbf{96.48 $\pm$ 0.13}  \\ \hline

	\end{tabular}%
	\label{tab:time-result}%
\end{table*}%

\begin{table*}[htbp]
	\setlength{\tabcolsep}{1.8mm}
	\centering
	\caption{The precision (90$\%$ observed links) of missing links prediction. \textbf{Bold} numbers are the best results of all methods.}
    \begin{tabular}{c|cccccccc}
      \hline
      \multirow{1}{*}{Data} & \multirow{1}{*}{Macaque} & \multirow{1}{*}{Mangwet}  & \multirow{1}{*}{Jazz} & \multirow{1}{*}{Metabolic} & \multirow{1}{*}{USAir} & \multirow{1}{*}{C.ele} & \multirow{1}{*}{E.coli} & \multirow{1}{*}{Yeast} \\ \hline

      \multirow{1}{*}{RA} & 0.5099 & 0.1292  & 0.5547 & 0.2451  & 0.4443 & 0.093 & 0.4857 & 0.2609 \\ \hline

      \multirow{1}{*}{CN} & 0.5695 & 0.125  & 0.5292 & 0.1127 & 0.3764 & 0.091 & 0.4399 & 0.1334  \\ \hline

      \multirow{1}{*}{LP} & 0.5483 & 0.1319  & 0.5109 & 0.1275 & 0.3821 & 0.089 & 0.4837 & 0.1454  \\ \hline

      \multirow{1}{*}{NMF} & 0.7316 & 0.4398  & 0.5309 & 0.2315 & 0.3981 & 0.1270 & 0.5013 & 0.3812  \\ \hline

      \multirow{1}{*}{RPCA} & 0.7421 & 0.5421  & 0.6138 & 0.1842 & 0.3596 & 0.098 & 0.3418 & 0.5359  \\ \hline

      \multirow{1}{*}{LFLP} & 0.7605 & 0.5572  & 0.5956 & 0.3241 & 0.4545 & 0.2010 & 0.5007 & 0.60  \\ \hline

      \multirow{1}{*}{GraphLP} & \textbf{0.7881} & \textbf{0.7986} & \textbf{0.8212} & \textbf{0.7030} & \textbf{0.8208} & \textbf{0.8224} & \textbf{0.7169} &  \textbf{0.7374}  \\ \hline

	\end{tabular}%
	\label{tab:time-result}%
\end{table*}%

\subsection{\label{sec:level1} Compared with Traditional Link Prediction Methods}

\begin{table*}[htbp]
	\setlength{\tabcolsep}{1.8mm}
	\centering
	\caption{The precision (90$\%$ observed links) of spurious links prediction. \textbf{Bold} numbers are the best results of all methods.}
    \begin{tabular}{c|cccccccc}
      \hline
      \multirow{1}{*}{Data} & \multirow{1}{*}{Macaque} & \multirow{1}{*}{Mangwet}  & \multirow{1}{*}{Jazz} & \multirow{1}{*}{Metabolic} & \multirow{1}{*}{USAir} & \multirow{1}{*}{C.ele} & \multirow{1}{*}{E.coli} & \multirow{1}{*}{Yeast} \\ \hline

      \multirow{1}{*}{RA} & 0.5490 & 0.1380  & 0.5410 & 0.140  & 0.2650 & 0.2790 & 0.4171 & 0.1110 \\ \hline

      \multirow{1}{*}{CN} & 0.5710 & 0.2880  & 0.5690 & 0.1670 & 0.2480 & 0.2458 & 0.3222 & 0.073  \\ \hline

      \multirow{1}{*}{LP} & 0.5939 & 0.3280  & 0.7016 & 0.6911 & 0.6271 & 0.4780 & 0.6089 & 0.4585  \\ \hline

      \multirow{1}{*}{NMF} & 0.8090 & 0.5660  & 0.6510 & 0.2430 & 0.4820 & 0.4333 & 0.4662 & 0.2330  \\ \hline

      \multirow{1}{*}{RPCA} & 0.810 & 0.5180  & 0.5920 & 0.074 & 0.443 & 0.2609 & 0.3795 & 0.4250  \\ \hline

      \multirow{1}{*}{LFLP} & 0.818 & 0.583  & 0.663 & 0.2210 & 0.5970 & 0.4390 & 0.4246 & 0.5680  \\ \hline

      \multirow{1}{*}{GraphLP} & \textbf{0.9073} & \textbf{0.8750} & \textbf{0.9197} & \textbf{0.8812} & \textbf{0.9057} & \textbf{0.9533} & \textbf{0.8452} &  \textbf{0.6210}  \\ \hline

	\end{tabular}%
	\label{tab:time-result}%
\end{table*}%

\begin{figure*}[!htb]
	\centering
	\includegraphics[scale=0.103]{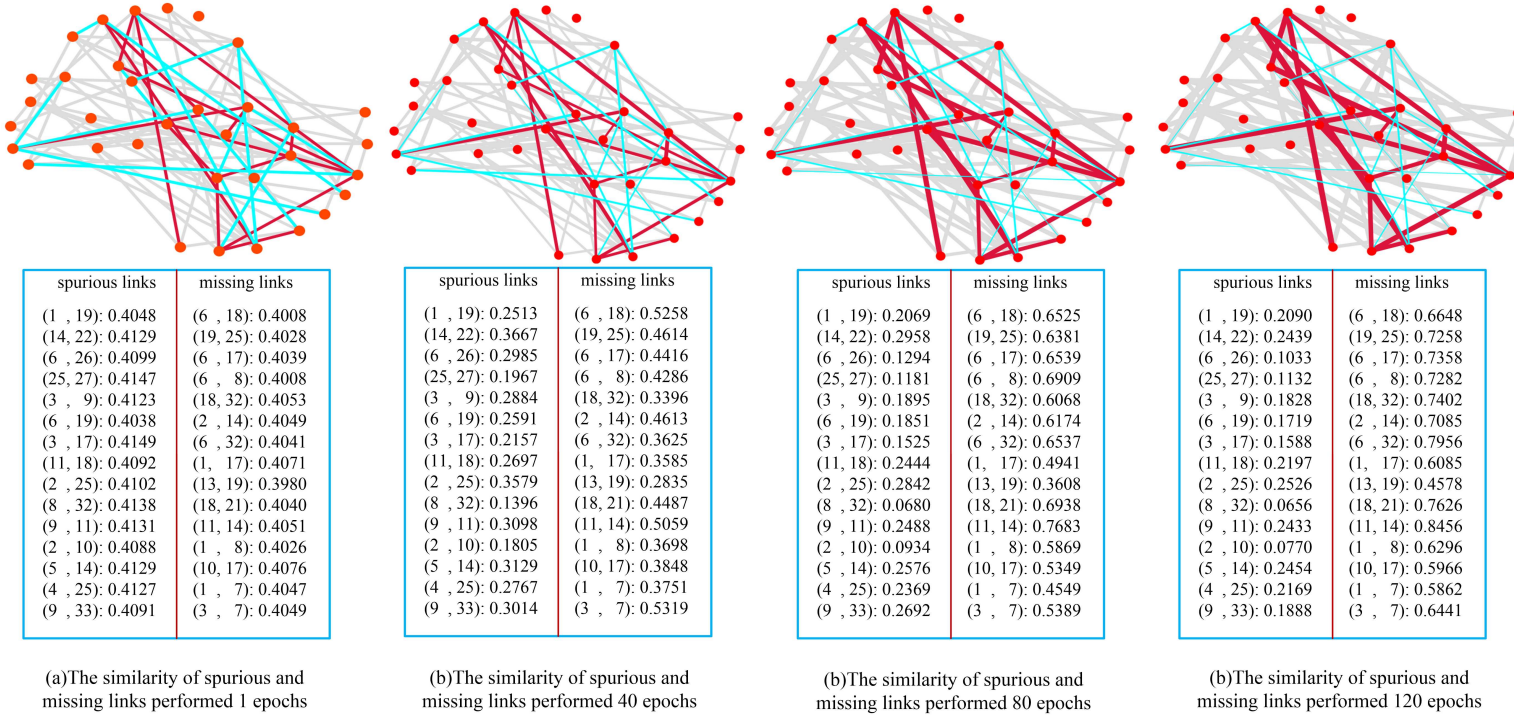}
	\caption{The topology visualization of Club dataset. The experiment performs 10\% link perturbation, i.e. 10\% spurious links are added and 10\% missing links are deleted.}
	\label{fig_sim:metattack}
\end{figure*}

\begin{figure*}[!htb]
	\centering
	\includegraphics[scale=0.074]{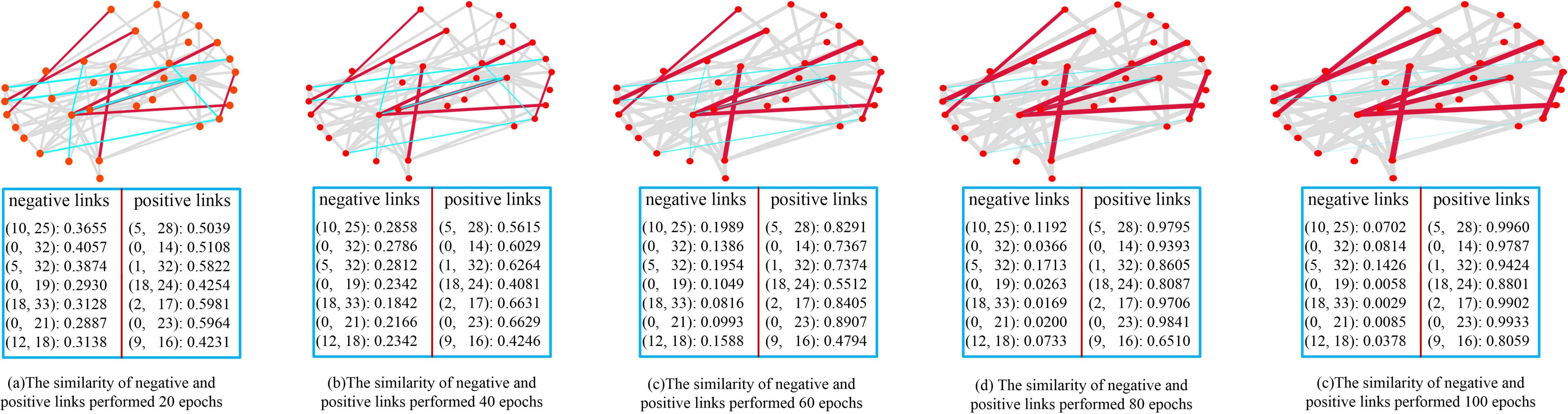}
	\caption{The topology visualization of Club dataset. The experiment performs 20\% link perturbation, i.e. 20\% spurious links are added and 20\% missing links are deleted.}
	\label{fig_sim:metattack}
\end{figure*}

To further verify the proposed method, the precision of GraphLP and traditional link prediction methods are calculated based on the following datasets: (1) Macaque~\citep{costa2007predicting},  cortical networks of the macaque monkey; (2) Mangwet~\citep{baird1998assessment}, the food web in Mangrove Estuary during the wet season; (3) Jazz~\cite{gleiser2003community}, a collaboration network of jazz musicians; (4) Metabolic~\citep{duch2005community}, a metabolic network of C.elegans; (5) USAir, (6) C.ele, (7) E.coli and (8) Yeast. Here, six representative traditional link prediction methods are selected for comparison.

(1) The CN~\citep{lorrain1971structural} metric is among the most widely used methods for link prediction problem. It assumes that two nodes will be more easily connected if they share more common neighbors;

(2) The RA~\citep{zhou2009predicting} metric is inspired by the physical processes involved in resource allocation, which suppresses the contribution of high-degree common neighbors;

(3) The LP~\citep{zhou2009predicting} index measures the structural similarity of node pairs within three-hops;

(4) The Non-negative Matrix Factorization (NMF)~\citep{lee2000algorithms} model is used for structure prediction by learning the latent features of real-world graphs;

(5) The Robust Principal Component Analysis (RPCA)~\citep{pech2017link} represents a real-world graph based on the sparsity and low rank property of its adjacency matrix and infers potential links based on matrix completion.


(6) The LFLP~\citep{xian2020netsre} uses a self-representation model to reconstruct the original graph based on a few representative subgraphs.

\begin{figure*}[tb]\centering
  \centering \subfigure[]
  { \includegraphics[width = 3.5in]{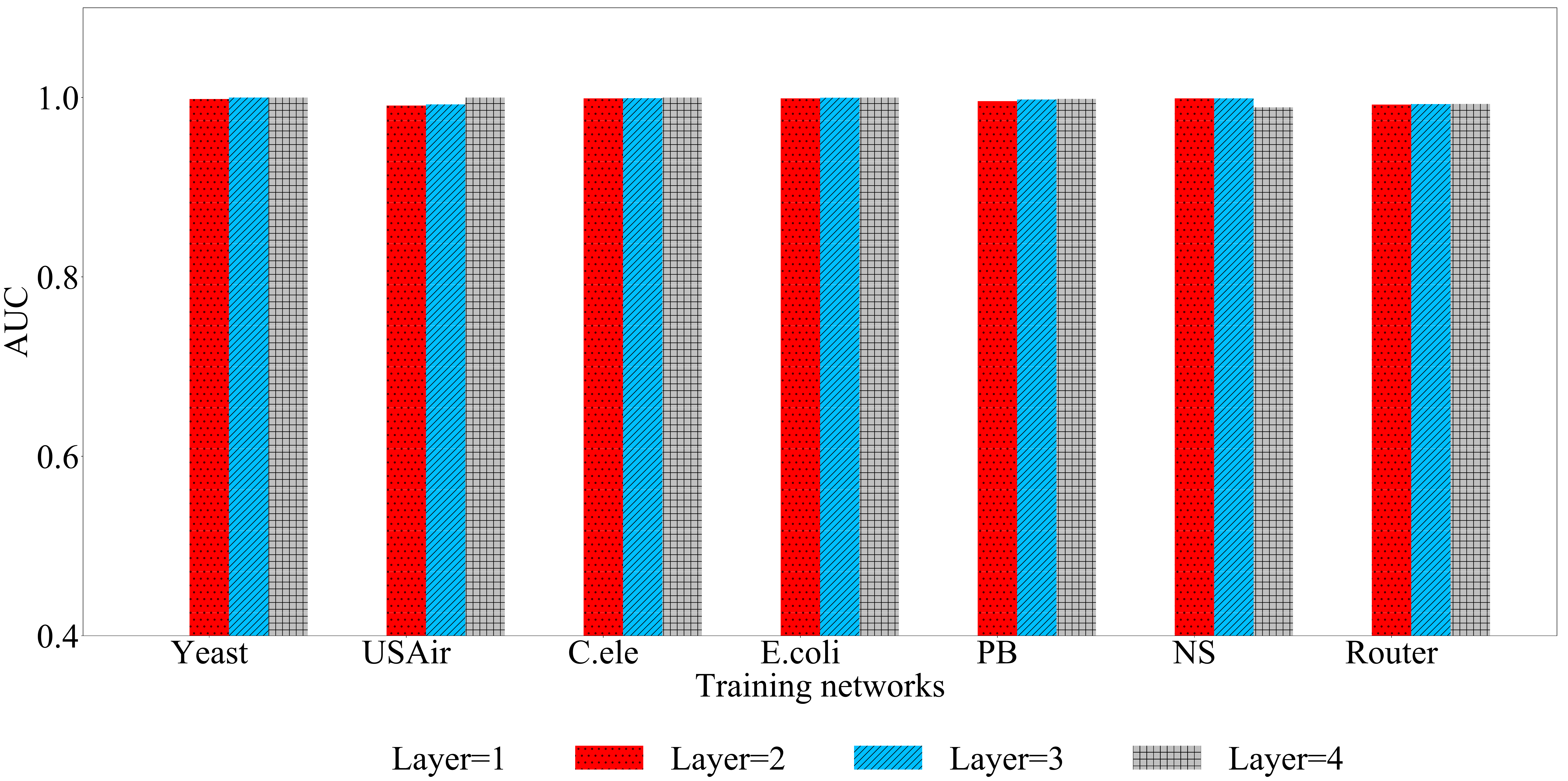}}
    \centering  \subfigure[]
  {\includegraphics[width = 3.5in]{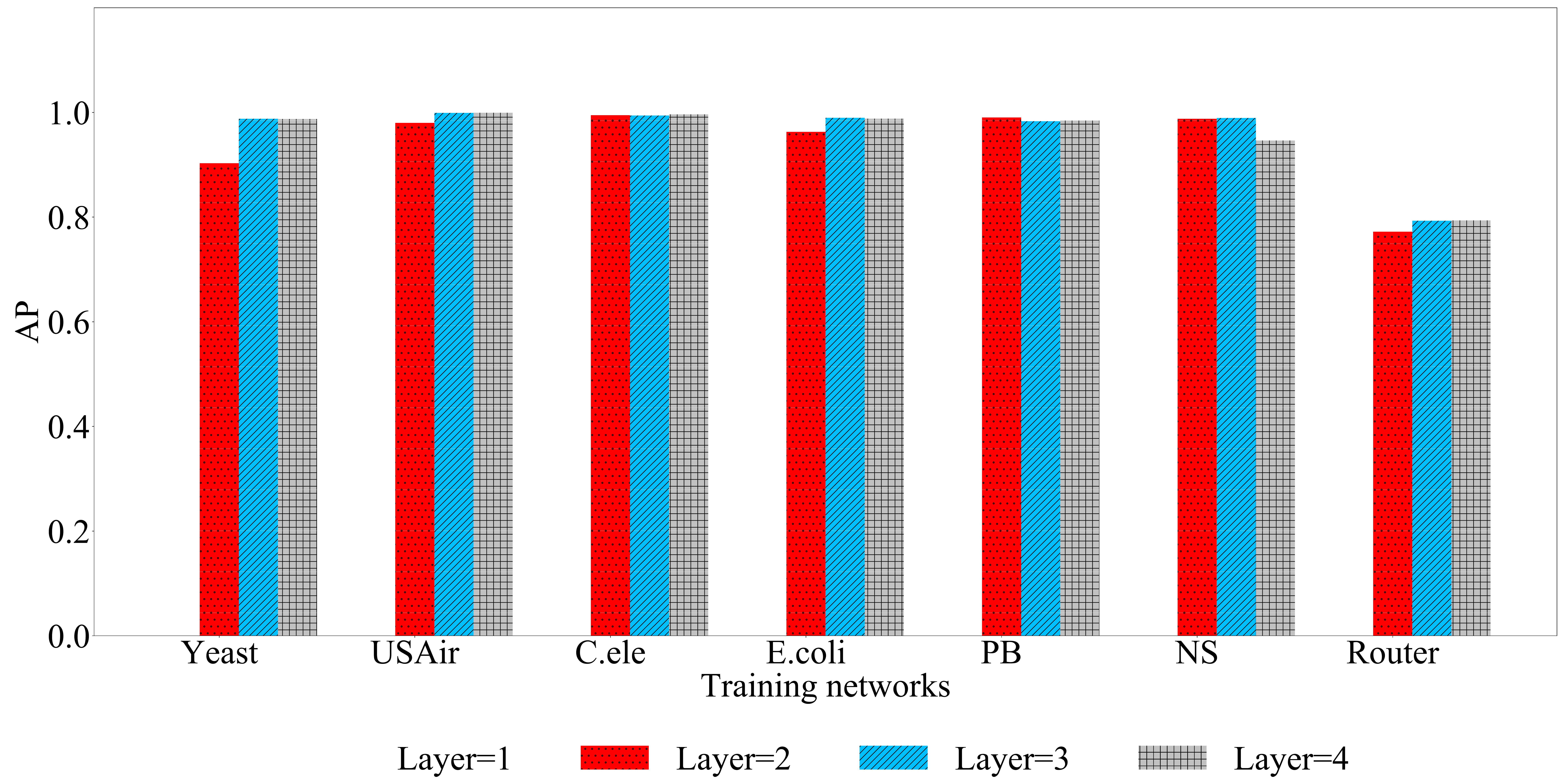}}
    \caption{The performance of link prediction under various model depth. (a) AUC of link prediction method GraphLP with different layer number. (b) AP of link prediction method GraphLP with different layer number.}
  \label{fig_sim:malicuous-edge}
\end{figure*}

\begin{figure*}[tb]\centering
  \centering \subfigure[]
  { \includegraphics[width = 3.5in]{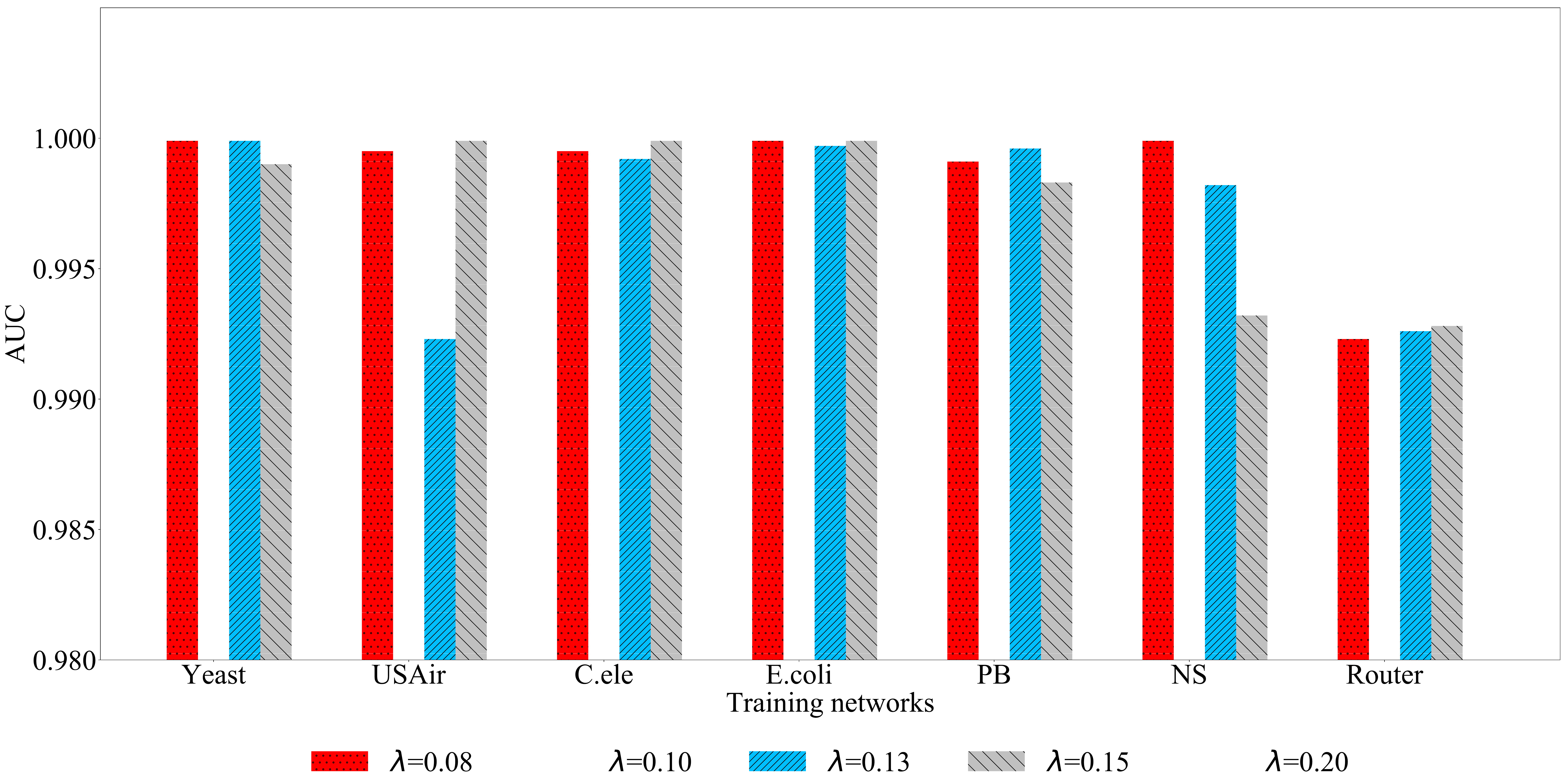}}
    \centering  \subfigure[]
  {\includegraphics[width = 3.5in]{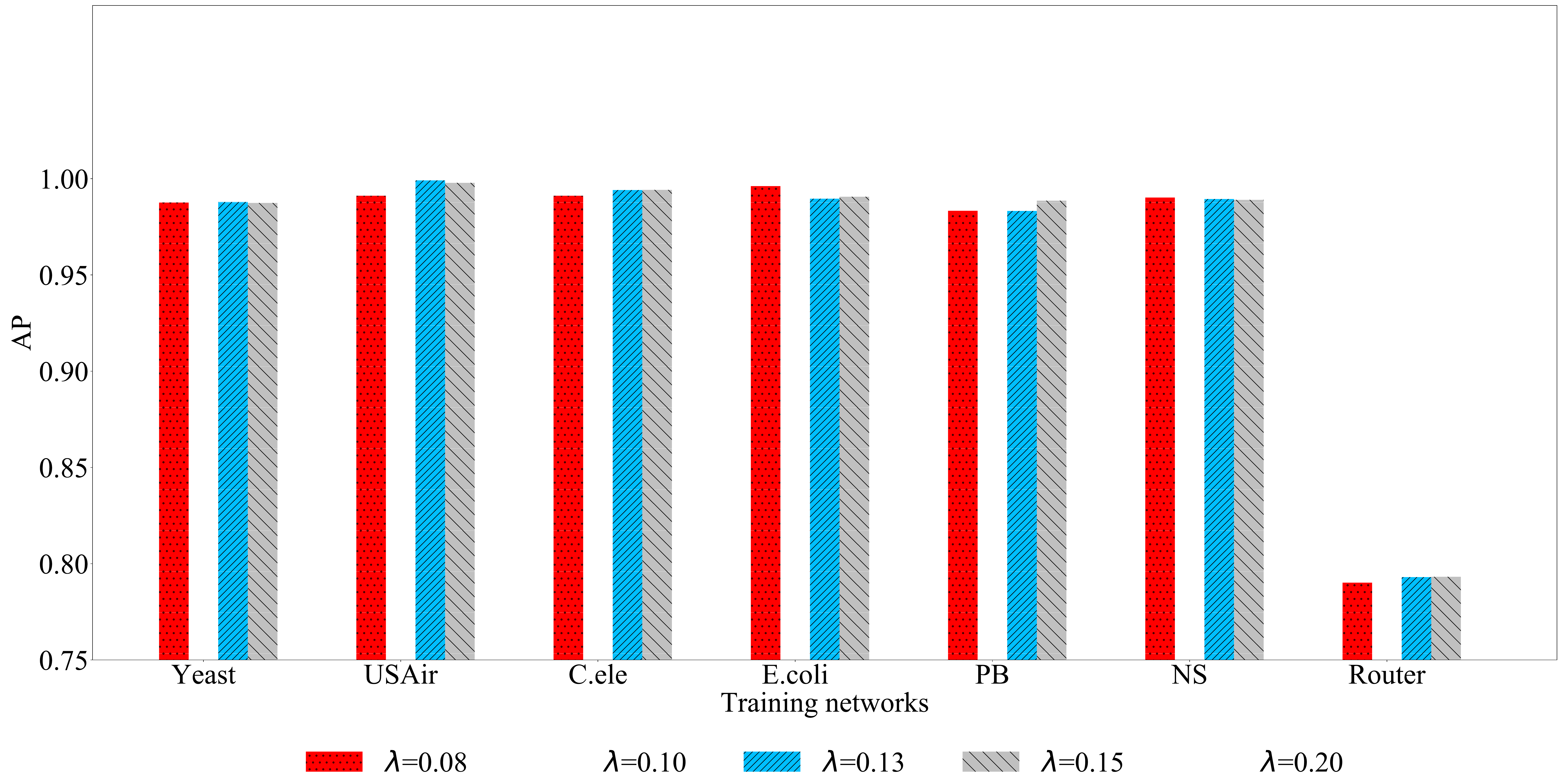}}
    \caption{The performance of link prediction under different values of $\lambda$. (a) AUC of link prediction method GraphLP under different values of $\lambda$. (b) AP of link prediction method GraphLP under different values of $\lambda$.}
  \label{fig_sim:malicuous-edge}
\end{figure*}



The results of missing link prediction with respect to Precision are shown in Table 8. For each network, the bold number in the corresponding column indicates the highest accuracy. The results presented in Table 8 demonstrate that the proposed model GraphLP model performs the best among the methods. Furthermore, the link prediction accuracy of the proposed model is far higher than that of the other methods, which can be at least three times higher than that of the best-performing method. For spurious links prediction, the results measured by Precision are listed in Table 9. For all networks, GraphLP performs the best among the methods and is remarkably better than the second best algorithm. The results presented in Table 8 and 9 demonstrate that GraphLP has stronger ability to learn structural features, and can recover the structure of the original network more accurately. Based on Table 2, it can be observed that the Precision of our proposed model performs best, despite the large differences between the ACC and AD across all the datasets; thus indicates that the proposed model performs well for heterogeneous graph structures.

\begin{figure*}[!tb]\centering
  \centering \subfigure[]
  { \includegraphics[width = 3.6in]{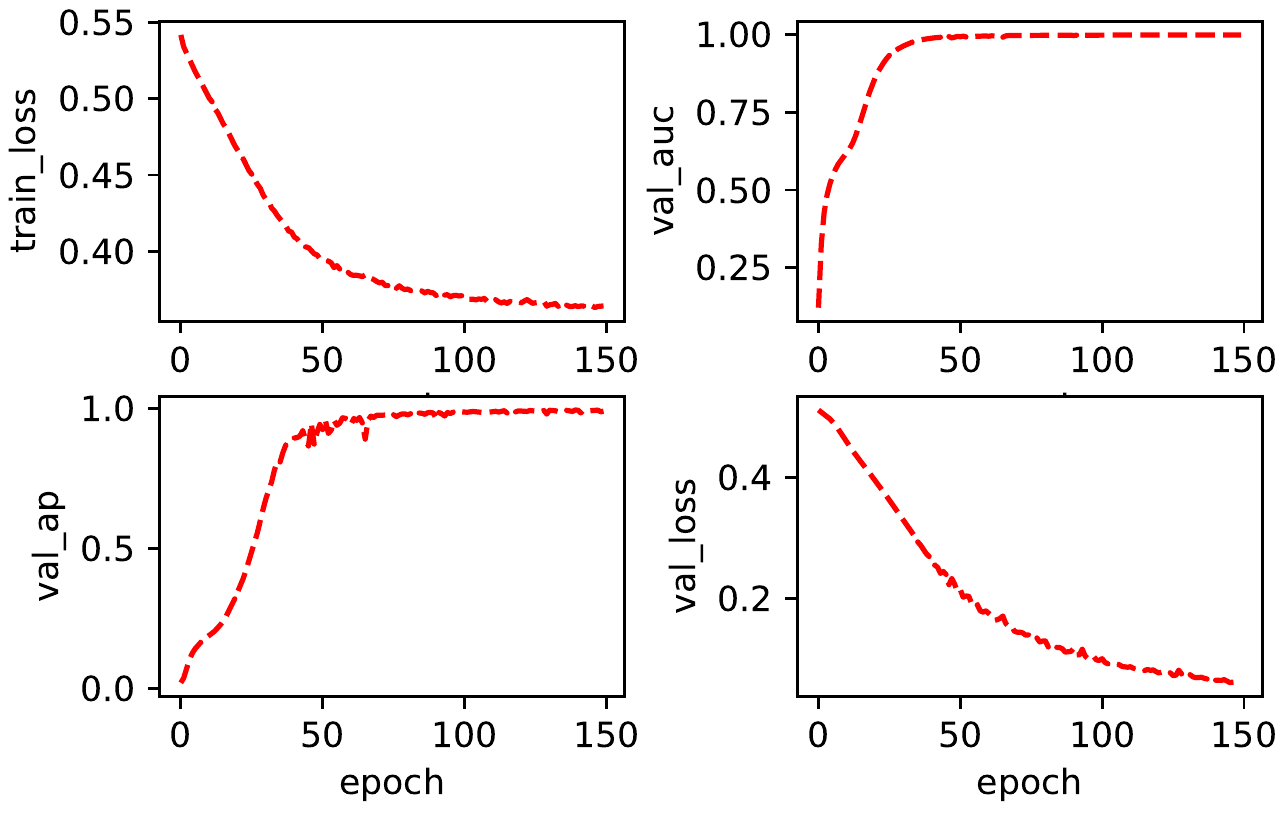}}
    \centering  \subfigure[]
  {\includegraphics[width = 3.5in]{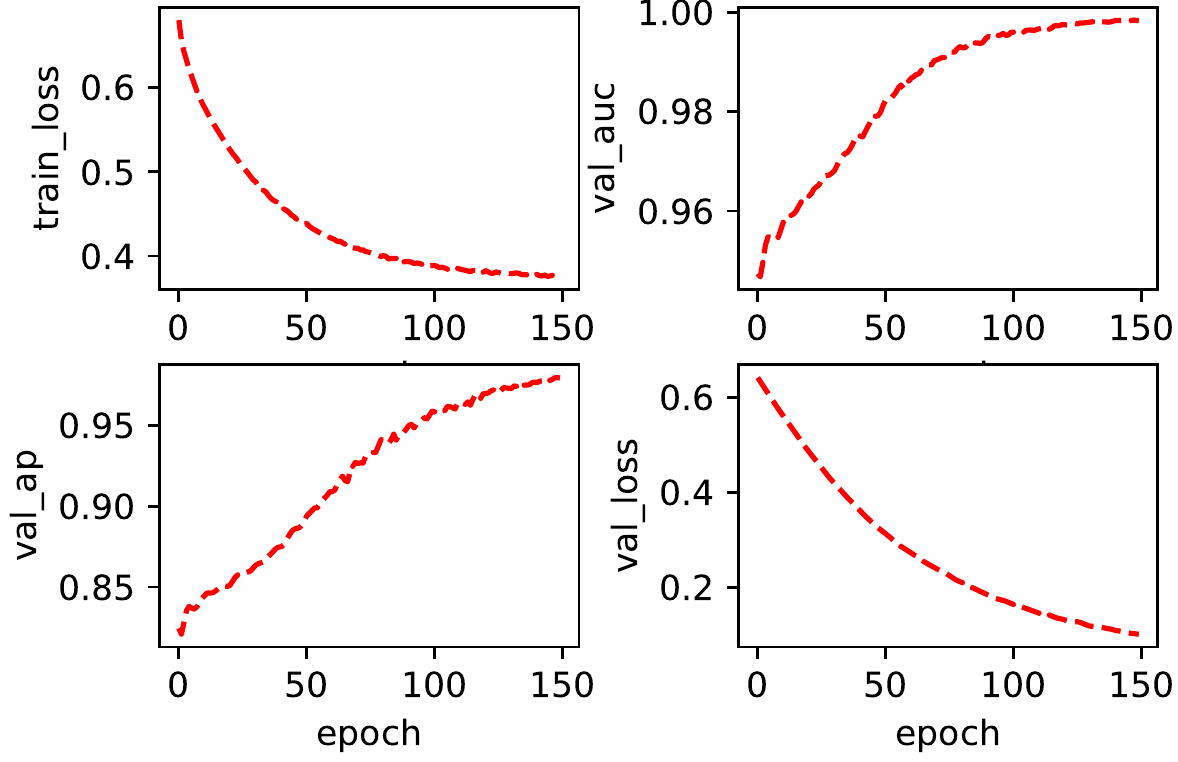}}
    \caption{Visualization of model convergence. (a) Convergence of USAir dataset. (b) Convergence of C.ele dataset.}
  \label{fig_sim:malicuous-edge}
\end{figure*}

\subsection{\label{sec:level1} Recovered Graph Visualization}
To verify the effectiveness of the proposed model of missing and spurious links inference, the topology of the recovered graphs in the model training process on Club dataset is visually compared, as depicted in Figure 4 and 5. The top half depicts the topology of the graphs, wherein the red links denote the missing links, the blue links denote the spurious links, and the gray links denote the original links. The bottom half depicts the likelihood scores of missing and spurious links. Based on the results, it can be concluded that with an increase in epoch times, the likelihood scores of missing links increases gradually, and the likelihood scores of spurious links decline gradually. The widths of the lines indicate the following process: when the number of epochs reaches 100, the likelihood scores of missing links approach 1.0, and the likelihood scores of spurious links approach 0.0. This proves that the proposed GraphLP model can distinguish between missing links and spurious links and infer them effectively. Moreover, to further prove the effectiveness of the proposed model, the topology of the recovered graphs in the model training process is visualized when 20\% of the links of Club are perturbed, as depicted in Figure 5. Compared to Figure 4, we determined that the likelihood of missing and spurious links is weakened with an increase in the structure perturbation ratio. However, with an increase in the training epochs, the model is still able to distinguish and infer the missing and spurious links according to the structural patterns. For instance, when the training epoch reaches 140, the likelihood scores of missing links are greater than 0.5, and those of spurious links are less than 0.3, indicating that the proposed model can still predict missing and spurious links with high accuracy.


\subsection{\label{sec:level1} Impact of Model Depth}
Next, the performance of GraphLP is explored at various model depths. As depicted in Figure 6, the performance of the model in terms of the AUC and AP trained with one-layer neural network is poor; however, its performance improves significantly with an increase in the number of layers. In particular, when the model depth is equal to two, a significant performance improvement is noted. The primary reason for this is that the model with two-layers multi-order global and local structural features is integrated adaptively based on the MLP component, which considerably improves the performance of the model. Subsequently, as the layer number increases, a slight improvement in model performance is still noted. When the number of layers is four, the accuracy of GraphLP declines significantly on NS and fluctuates on other datasets. A possible reason for this is that the model with four layers becomes more complex, thereby requiring more training iterations or an appropriate learning rate \citep{cong2021provable}. In general, the performance of the proposed model is optimal when the depth is three, and a deep architecture is necessary.

\subsection{\label{sec:level1} Impact of Trade-off Parameter}
To examine the sensitivity of the proposed model to the trade-off parameter, the AUC and AP values of link prediction methods with different $\lambda$ are presented in Figure 7. Based on the results, it can be concluded that the performance of the proposed model is not sensitive to $\lambda$ for most datasets. In Figure 7(a), for the USAir and NS datsets, the AUC value varies significantly under different $\lambda$, but the performance is still better than other of the other algorithms. In Figure 7(b), the AP value remains stable for different $\lambda$ values, indicating that the proposed model is insensitive to different $\lambda$. Overall, our proposed algorithm exhibited satisfactory performance on most datasets with various $\lambda$.

\subsection{\label{sec:level1} The Convergence Analysis}
Generally, GraphLP converges to optimal values after approximitely 200 epochs on most datasets. In particular, Figure 8 plots the learning curves of GraphLP on the USAir and C.ele datasets, including the training loss, validation AUC, validation AP, and validation loss. The results indicate that the AUC and AP values increase rapidly with the decrease in training loss and validation loss, and these values converge to the optimal value when the validation loss approaches a minimum value. Additionally, we discover that validation loss is lower than training loss, and the difference between them remains relatively stable. A possible reason for this is that the dropout manipulation is only applied to the training process.



\section{Conclusion}
This paper aims to reconstruct graph structure to improve the performance of link prediction. In particular, unlike existing subgraph-classification-based discriminative methods, this work achieves the aforementioned objective by developing a generative GNN, namely GraphLP, which considered both global and local structure features and hierarchical structural patterns. Concurrently, a novel collaborative inference operation and high-order connectivity computation mechanism are developed. We also present an analysis about the relation between GraphLP and other classical link prediction methods. Extensive experimental results demonstrate the superiority of the proposed method over other state-of-the-art models and traditional baseline methods. This could be a fruitful avenue for future research aimed at addressing graph learning tasks.

\section*{Acknowledgment}

This work was partially supported by the National Natural Science Foundation of China under Grant Nos. 62106030, 61802039, 62272066; Chongqing Municipal Postdoctoral Science Foundation under Grant No. cstc2021jcyj-bsh0176; Chongqing Municipal Natural Science Foundation under Grant No. cstc2020jcyj-msxmX0804; the Chongqing Research Program of Basic Research and Frontier Technology under Grant No. cstc2021jcyj-msxmX0530.

\bibliographystyle{plainnat}
\bibliography{GraphLP}

\end{document}